\documentclass[twocolumn]{aastex62}

\usepackage{amsmath}

\def \hbeta{H$\beta$}
\def \ebmv{E(B-V)}
\def \ebmvs{ E_{s}{\rm (B-V)} }
\def \ebmvn{ E_{n}{\rm (B-V)} }
\def \halpha{H$\alpha$}
\def \Msol{{M}_{\odot}}

\def \Zsol{{\rm Z}_{\odot}}

\def \logm{\log(M/\Msol)}
\def \lya{Ly$\alpha$}
\def \ewha{{\rm EW}({\rm H}\alpha)}

\def \h2{{\rm H_{2}}}

\def \hbeta{H$\beta$}
\def \halpha{H$\alpha$}
\def \siii{Si{\scriptsize ~II}}

\def \niiha{[N{\scriptsize ~II}]/H$\alpha$}

\def \nii{[N{\tiny\,II}]}

\def \LUV{L_{{\rm UV}}}
\def \LHA{L_{{\rm H\alpha}}}
\def \DLHA{\Delta\log(L_{{\rm H\alpha}})}

\def \dn4000{D_{{\rm n}}(4000) }
\def \flya{$f_{\rm Ly\alpha}$}
\def \iracA{[$3.6\,{\rm \mu m}$]$-$[$4.5\,{\rm \mu m}$]}

\def \fion{$f^{\rm ion}_{\rm esc}$}
\def \xiion{$\xi_{\rm ion}$}

\graphicspath{{./}{figures/}}

\accepted{for publication in ApJ, September 2019}
\shorttitle{Recent Burstiness Star Formation in Galaxies at $z\sim4.5$}
\shortauthors{Faisst et al.}
\begin{document}

\title{\sc \large The Recent Burstiness of Star Formation in Galaxies at $z\sim4.5$ from \halpha~Measurements}

\correspondingauthor{Andreas L. Faisst}
\email{afaisst@caltech.edu}

\author[0000-0002-9382-9832]{Andreas L. Faisst}
\affil{IPAC, California Institute of Technology, 1200 East California Boulevard, Pasadena, CA 91125, USA}

\author[0000-0003-3578-6843]{Peter L. Capak}
\affil{IPAC, California Institute of Technology, 1200 East California Boulevard, Pasadena, CA 91125, USA}
\affil{Cosmic Dawn Center (DAWN), Copenhagen, Denmark}

\author[0000-0003-2047-1689]{Najmeh Emami}
\affil{Department of Physics and Astronomy, University of California Riverside, Riverside, CA 92521 USA}

\author[0000-0002-8224-4505]{Sandro Tacchella}
\affil{Center for Astrophysics $\mid$ Harvard \& Smithsonian, 60 Garden St, Cambridge, MA 02138, USA}

\author[0000-0003-3917-6460]{Kirsten L. Larson}
\affil{Department  of  Astronomy,  California  Institute  of  Technology,  1200  E. California  Blvd.,  Pasadena,  CA  91125,  USA}

%\author{others?}
%\affil{}



\begin{abstract}

The redshift range $z=4-6$ marks a transition phase between primordial and mature galaxy formation in which galaxies considerably increase their stellar mass, metallicity, and dust content. The study of galaxies in this redshift range is therefore important to understand early galaxy formation and the fate of galaxies at later times. Here, we investigate the burstiness of the recent star-formation history (SFH) of $221$ $z\sim4.5$ main-sequence galaxies at $\logm>9.7$ by comparing their ultra-violet (UV) continuum, \halpha~luminosity, and \halpha~equivalent-width (EW). The \halpha~properties are derived from the Spitzer \iracA~broad-band color, thereby properly taking into account model and photometric uncertainties.
We find a significant scatter between \halpha~and UV-derived luminosities and star-formation rates (SFRs). About half of the galaxies show a significant excess in \halpha~compared to expectations from a constant smooth SFH. We also find a tentative anti-correlation between \halpha~EW and stellar mass, ranging from $1000\,{\rm \AA}$ at $\logm < 10$ to below $100\,{\rm \AA}$ at $\logm > 11$.
Consulting models suggests that most $z\sim4.5$ galaxies had a burst of star-formation within the last $50\,{\rm Myrs}$, increasing their SFRs by a factor of $>5$. The most massive galaxies on the other hand might decrease their SFRs, and may be transitioning to a quiescent stage by $z=4$.
We identify differential dust attenuation ($f$) between stars and nebular regions as the main contributor to the uncertainty. With local galaxies selected by increasing \halpha~EW (reaching values similar to high-$z$ galaxies), we predict that $f$ approaches unity at $z>4$ consistent with the extrapolation of measurements out to $z=2$.

\end{abstract}



%%%%%%%%%%%%%%%%%%%%%%%%%%%%%
%%% 	INTRODUCTION 	  %%%
%%%%%%%%%%%%%%%%%%%%%%%%%%%%%%
\section{Introduction} \label{sec:intro}

The question of how the most early galaxies formed, how they evolved, and how they became like our Milky Way or the rest of modern galaxies is one of the main drivers of current extragalactic research. The redshift range $z=4-6$ ($0.9-1.5$ billion years after the Big Bang) is of particular interest as galaxies undergo substantial changes in their morphology, star-formation activity, and chemical composition.

Large campaigns with the Hubble Space Telescope (HST) have led to large samples of galaxies at $z>4$ in numerous fields on sky \citep[e.g.,][]{TRENTI11,FINKELSTEIN12,BOUWENS15,ONO18}. These led to a better understanding of their rest-frame ultra-violet (UV) light and in particular the time-evolution of the cosmic star-formation rate (SFR) density \citep[e.g.,][]{BOUWENS14,OESCH18}.
In parallel, ground-based narrow-band imaging surveys added to the population diversity by identifying hundreds of young, star-forming \lya~emitters at discrete redshifts above $z=4$ \citep{MATTHEE17,KONNO18,OUCHI18,SHIBUYA18a}.
Large programs with ground-based optical spectrographs verified the redshifts of a large fraction of high-redshift candidate galaxies, to provide robust samples of hundreds of galaxies beyond $z=4$ \citep{VANZELLA07,KASHIKAWA11,VANDOKKUM13,LEFEVRE15,HASINGER18}.
The Spitzer Space Telescope is crucial to study $z>4$ galaxies as it is currently the only telescope able to observe the rest-frame optical light of these galaxies. It provides measurements of stellar ages and masses and optical emission lines \citep[e.g.,][]{SCHAERER09,SHIM11,LABBE13,STARK13,GONZALEZLOPEZ14,SMIT14,STEINHARDT14,OESCH15,BOUWENS16b,FAISST16b,FAISST16a,RASAPPU16,DAVIDZON17,HARIKANE18}.

All these studies show that galaxies significantly change their chemical and morphological properties of their interstellar medium (ISM) across cosmic time. For example, in only $600\,{\rm Myrs}$ between redshifts $z=6$ and $4$, they substantially increase their stellar mass, dust, and metal content \citep{BOUWENS09,ILBERT13,MADAU14,FAISST16b,DAVIDZON17}, while producing stars at rates that are more than a factor of $50$ higher compared to the present time \citep{LILLY13,FELDMANN15,TASCA15,RASAPPU16,FAISST16a,DAVIDZON18}. The latter is likely connected to larger gas reservoirs and higher star formation efficiency at early times \citep{TACCONI10,TACCONI13,GENZEL15,SILVERMAN15,SCOVILLE16}. This is also suggested by large offsets between UV and far-infrared continuum emission that might be caused by the strong radiation pressure and stellar winds of young star-forming regions and the interaction between galaxies in a dense early universe \citep[][]{MAIOLINO15,FAISST17b}.
In addition to mostly young galaxies, several massive ($\logm \gtrsim 11$) galaxy candidates have been found beyond redshift of $z=4$ out to $z=6$ \citep{STEINHARDT14}. These must have gained their masses during only a couple of $100\,{\rm Myrs}$ and therefore put constraints on models of galaxy formation in a cosmological context \citep{TACCHELLA13,STEINHARDT14,BEHROOZI18,TACCHELLA18,VOGELSBERGER19} as well as the population of the first quiescent galaxies at $z>3$ \citep{GOBAT12,STRAATMAN14,GLAZEBROOK17,KUBO18,SCHREIBER18,GUARNIERI19}.
This population diversity made out of young pristine and more evolved galaxies is also reflected in studies of UV absorption lines that suggest large range in the amount of metals \citep[][]{ANDO07,FAISST16b}. This is backed up by ALMA observations indicating a large range of dust masses in the same population of galaxies \citep[][]{CAPAK15,FAISST17b,FUDAMOTO17,FAISST19}.

Constraints on the recent star-formation history (SFH) can reveal more details about the formation of these galaxies.
Studies at low redshifts suggest differences in the star-formation activity of low- and high-mass galaxies. Low-mass and dwarf galaxies ($\logm < 9$) display a bursty SFH that is likely caused by feedback from AGN and supernovae events \citep{SULLIVAN01,IGLESIASPARAMO04,BOSELLI09,WUYTS11,WEISZ12,DOMINGUEZ15,GUO16,SHIVAEI16,IZOTOV17,SPARRE17,EMAMI18,BROUSSARD19,CAPLAR19}. The same studies show that more massive galaxies ($\logm \sim10$) have smoother SFHs, with variations on time scales of several $100\,{\rm Myrs}$, for example caused by changes in the gas inflow rates \citep[e.g.,][]{WEISZ12,SPARRE15,TACCHELLA16,EMAMI18}.
Not much is known about the burstiness of the SFH at early cosmic times. From the various detailed studies at UV and FIR wavelengths, we know that the ISM of high-$z$ galaxies is complex $-$ influenced by internal physics, high gas fractions, gas in- and outflows, and galaxy interactions. A bursty SFH is therefore expected, similar to local dwarf galaxies, although the causes of burstiness may be different at early cosmic times \citep[][]{FAUCHER18}. 
Furthermore, the study of the SFH of very massive galaxies at $\logm > 11$ might reveal something about the creation of the first quiescent galaxies.  
	
%%%%%% FIGURE: LINES %%%%%%%
\begin{figure}
\includegraphics[width=1.0\columnwidth, angle=0]{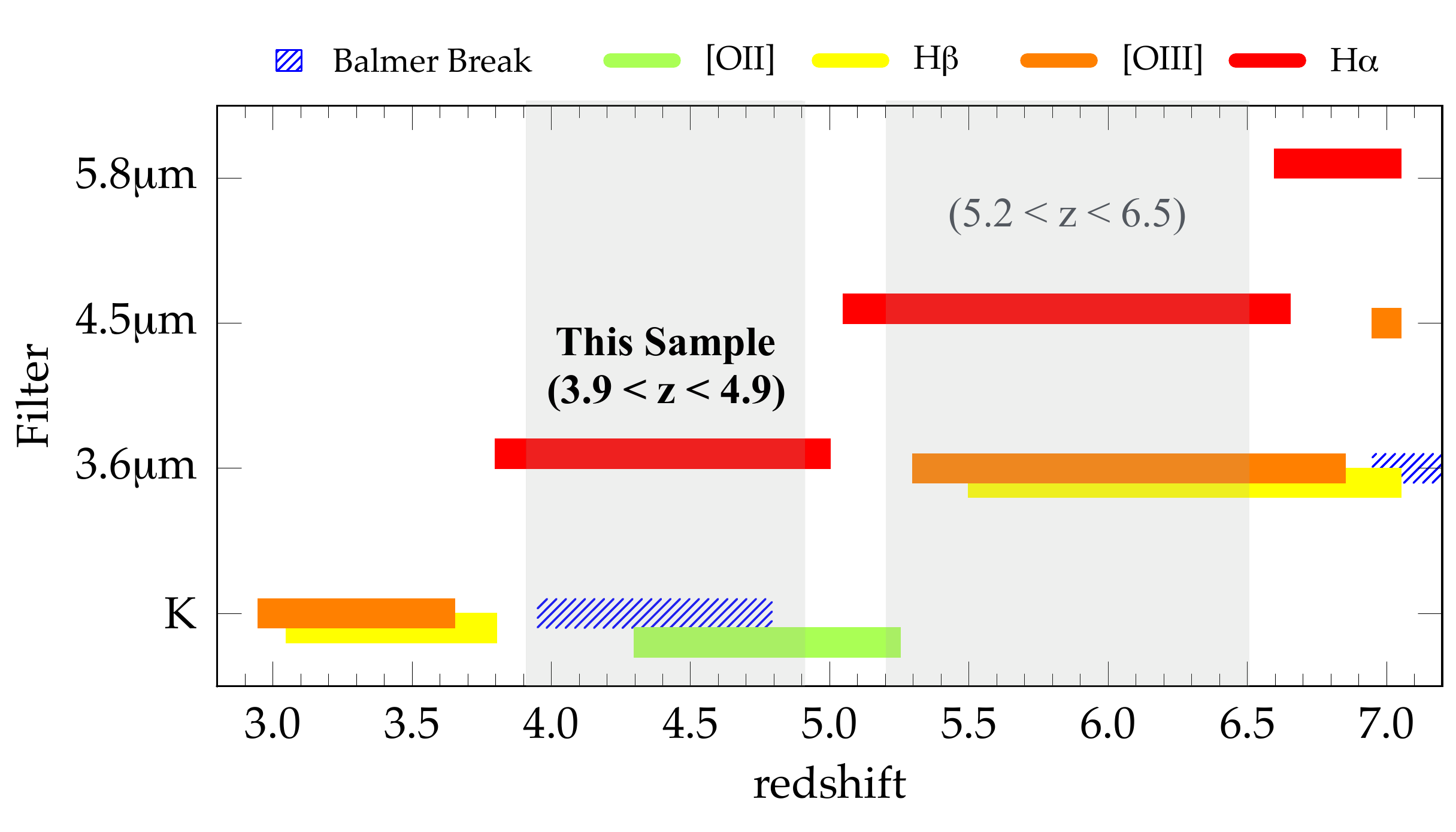}
\caption{Contribution of strong optical emission lines to various broad-band filters as a function of redshift. Here, we focus on the redshift range $3.9 < z < 4.9$ where we can access the \halpha~luminosities using the Spitzer broad-band filters at $3.6\,{\rm \mu m}$ and $4.5\,{\rm \mu m}$ without contamination by other strong emission lines. \label{fig:lines}}
\end{figure}
%%%%%%%%%%%%%%%%%%%%%%%%%%%%%
	
While at low redshifts a combination of photometry and high-resolution spectroscopy can be used to reconstruct the SFH of galaxies \citep{CONROY16,IYER17,CARNALL18,LEJA18}, it is more difficult at $z>4$ due to the current lack of high-resolution spectroscopic coverage at rest-frame optical and near-IR wavelengths.
The comparison of the light emitted from \halpha~and UV continuum can provide constraints on the recent star-formation activity of galaxies. While the former is emitted by recombination in nebular clouds primarily due to $O$-type stars on timescales of a few $10$s of Myrs, the latter is produced by $O$- and $B$-type stars with lifetimes of $\sim100\,{\rm Myrs}$ \citep[e.g.,][]{KENNICUTT98}. Hence, the two tracers couple to star formation on different timescales, and provide constraints on the recent SFH and its burstiness. Several studies at $z<3$ have shown the feasibility of this method \citep[e.g.,][]{WEISZ12,EMAMI18,CAPLAR19}.

In this work, we put constraints on the recent SFHs and burstiness of star-formation activity of $221$ galaxies at $z\sim4.5$ spanning a large range of stellar masses. For this, we combine measurements of their rest-frame UV continuum luminosity with that of \halpha~line emission. The latter is derived from Spitzer colors at $3.6$ and $4.5\,{\rm \mu m}$ \citep[Figure~\ref{fig:lines} and][]{FAISST16a}, as spectroscopic observations at these wavelengths are currently not possible for galaxies at these redshifts.

The outline of this paper is as follows.
In Section~\ref{sec:data}, we introduce the dataset and the sample selection. In Section~\ref{sec:emlinfit}, we detail our method to derive \halpha~luminosities for individual galaxies at $z>4$ from the Spitzer color and subsequently evaluate our technique by performing various simulations and tests. The results of our measurements (Section~\ref{sec:results}) are discussed in connection with theoretical models in Section~\ref{sec:discussion}. We conclude and summarize in Section~\ref{sec:end}.
Throughout this work, we assume a flat cosmology with $\Omega_{\Lambda,0}=0.7$, $\Omega_{m,0}=0.3$, and $h = 0.7$. Stellar masses and SFRs are scaled to a \citet{CHABRIER03} initial mass function (IMF) and magnitudes are quoted in the AB system \citep{OKE74}.

%%%%%%%%%%%%%%%%%%%%%%%%%%%%%
%%%		DATA     		  %%%
%%%%%%%%%%%%%%%%%%%%%%%%%%%%%
\section{Data} \label{sec:data}

\subsection{Photometric and Spectroscopic Datasets}\label{sec:mainsampleselection}

Our sample is selected from exquisite photometric and spectroscopic redshift data available on the \textit{Cosmic Evolution Survey} field \citep[COSMOS\footnote{\url{http://cosmos.astro.caltech.edu}},][]{SCOVILLE07}.
Notably, COSMOS is covered by deep Spitzer imaging from the \textit{Spitzer Large Area Survey with Hyper Suprime-Cam} \citep[SPLASH\footnote{\url{http://splash.caltech.edu}},][]{STEINHARDT14,LAIGLE16}, whose depth at $3.6\,{\rm \mu m}$ and $4.5\,{\rm \mu m}$ enables a robust measurement of \halpha~emission for a large sample of galaxies. COSMOS has been targeted by many spectroscopic surveys over the past years. This led to large spectroscopic samples of galaxies at $z>4$ confirmed by \lya~emission as well as strong rest-frame UV absorption features. Such a mixed selection reduces typical biases of spectroscopic samples towards highly star-forming and dust-poor galaxies \citep[e.g.,][]{BARISIC17}.
Our main sample is selected from photometric and spectroscopic data at $3.9 < z < 4.9$, where the Spitzer color can be used to constrain \halpha~emission (see Figure~\ref{fig:lines}).

\textbf{The photometric sample} is based on the photometric redshifts provided in the \textit{COSMOS2015} catalog \citep{LAIGLE16}, which are accurate enough ($\sim3\%$ at $z>4$) for selecting galaxies in this narrow redshift window \citep[see][]{FAISST16a}. We only consider redshift probability distribution functions (PDFs) with no secondary (low-redshift) solution at a significance level greater than $5\%$ and a $1\sigma$ redshift uncertainty\footnote{The $1\sigma$ width is defined as $|z_{l} - z_{u}|/(1+z)$, where $z_l$ and $z_u$ are the lower and upper $1\,\sigma$ limits of the photometric redshift PDF for a given most likely redshift $z$.} of less than $2\%$.

\textbf{The spectroscopic sample} is based on the COSMOS spectroscopic master catalog (Salvato et al., in prep.), which includes all spectroscopic observations carried out on COSMOS to-date. 
The main contribution to the redshifts at $z>4$ comes from the VIMOS Ultra Deep Survey \citep[VUDS,][]{LEFEVRE15,TASCA17} and the Deep Imaging Multi-Object Spectrograph (DEIMOS) survey program at Keck \citep{HASINGER18}. We select robust spectroscopic redshifts by using the quality flags provided in the catalog. Specifically, we use flags 3, 4, 13, 14, 23, 24, 18, 214, 213, which encompass the most secure redshift measurement for the targeted galaxies as well as serendipitous detections and at the same time remove optically bright AGN.

\subsection{Final Galaxy Sample}

We apply additional cuts to the redshift selected samples to ensure the robust measurement of the \halpha~luminosity and in the following quantify the completeness of the final sample.

\subsubsection{Signal-to-Noise Selection}\label{sec:snselection}
We focus on the brightest and most massive (see Section~\ref{sec:completeness}) galaxies in order to be able to measure \halpha~luminosity and EW at high confidence.
Specifically, we require a signal-to-noise (S/N) ratio of $10$ at $3.6\,{\rm \mu m}$ and $5$ at $4.5\,{\rm \mu m}$\footnote{Note that the galaxies are generally brighter at $3.6\,{\rm \mu m}$ due to the \halpha~emission (unless they are dust obscured), hence the same S/N cut in both bands would result in biases against strong \halpha~emitters at the survey limit.}.
As shown in detail in Section~\ref{sec:emlinfit}, this will enable the measurement of the \halpha~luminosity within a factor of two uncertainty for our final galaxy sample given the survey flux limits of the \textit{SPLASH} data \citep[$\sim25.5\,{\rm AB}$ at $3\sigma$,][]{LAIGLE16}.
Applying this S/N cut to our full sample (Section~\ref{sec:mainsampleselection}), we end up with $294$ ($165$) galaxies with spectroscopic (photometric) redshifts at $3.9 < z < 4.9$.

%%%%%% FIGURE: SAMPLE SUMMARY 2 %%%%%%%
\begin{figure}
\includegraphics[width=1.0\columnwidth, angle=0]{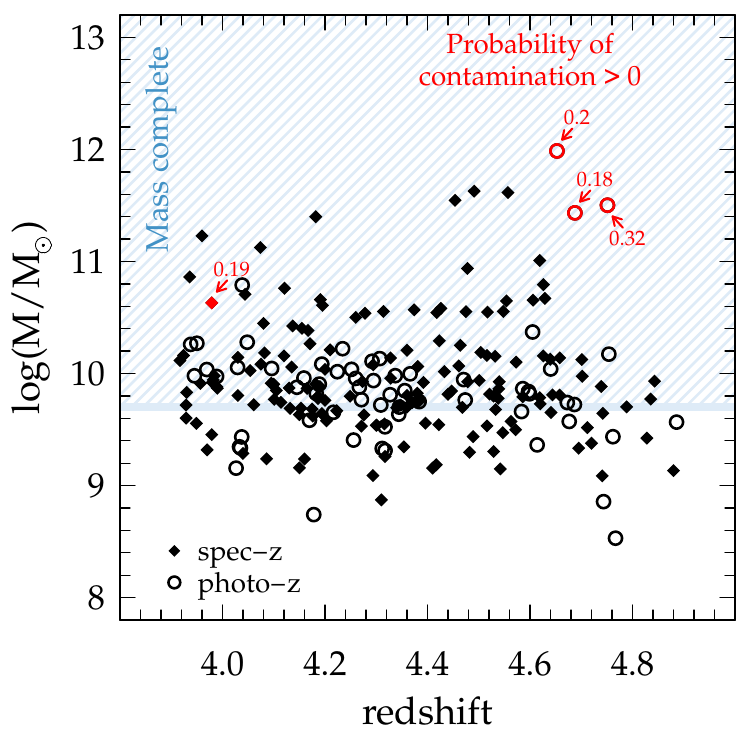}
\caption{Stellar mass distribution as a function of redshift for the spectroscopic (filled diamonds) and photometric (open circles) sample. The red arrows and numbers indicate galaxies with $\logm > 10.5$ whose Spitzer photometry is contaminated with a non-zero probability according to our machine-learning method (see Appendix~\ref{sec:appendixA}). These four galaxies are removed from the final sample. \label{fig:sample}}
\end{figure}
%%%%%%%%%%%%%%%%%%%%%%%%%%%%%

\subsubsection{Selection by Robust Spitzer photometry}\label{sec:spitzercontamination}
The large ($2.5\arcsec$) full-width-at-half-maximum (FWHM) of the Spitzer/IRAC point spread function (PSF) at $4.5\,{\rm \mu m}$ causes blending of nearby sources. This hampers a robust determination of their photometry and consequently their \halpha~emission.
We therefore remove galaxies with potentially contaminated Spitzer photometry. For this, available ACS/F814W ($\sim8000\,{\rm \AA}$) imaging from the Hubble Space Telescope \citep{KOEKEMOER07,SCOVILLE07b} as well as the $K$-band imaging from UltraVISTA \citep{MCCRACKEN12,LAIGLE16} is used.
First, we remove galaxies with more than one detection (including themselves) in F814W within a radius of half the FWHM of the Spitzer PSF at $4.5\,{\rm \mu m}$ ($175$ galaxies, $38\%$). This leads to a reduced sample of $284$ galaxies.
Four of these galaxies do not have ACS imaging and $16$ are not detected in F814W, likely due to significant dust obscuration.
In a final step, we perform a visual inspection of the remaining galaxies (including the $20$ galaxies without ACS data or detection) using the ground-based $K$-band images from UltraVISTA \citep[median seeing FWHM of $0.7\arcsec$;][]{FAISST17a}. Due to several dusty, F814W-undetected foreground galaxies in close projected distance, we remove another $22\%$ of the galaxies. We end up with $221$ galaxies that represent our final sample.

In Appendix~\ref{sec:appendixA}, we describe a more quantitative way of assessing the contamination of the Spitzer photometry by using unsupervised machine learning. While there is not much gain in our case, this technique is powerful if applied to (much larger) future datasets. In brief, the \texttt{t-SNE} algorithm \citep{MAATEN08} that used here, groups similar configurations of galaxies on an image basis. This allows us to derive the probability at which a given galaxy's photometry is contaminated. Here we convert this probability into a true/false binary number by defining contamination by a probability of $\geq0.5$. With this definition and by comparing to our visual classification (which we take as the truth), we find a purity of $75\%$ and completeness of $70\%$. These numbers are likely to improve with larger sample sizes.

%%%%%% FIGURE: SAMPLE SUMMARY 1 %%%%%%%
\begin{figure*}
\includegraphics[width=2.0\columnwidth, angle=0]{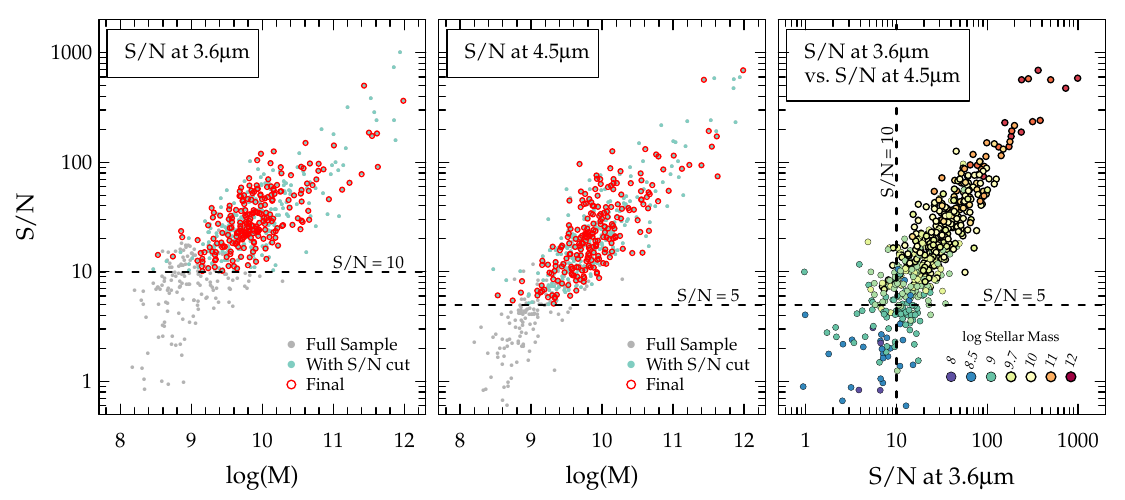}
\includegraphics[width=2.0\columnwidth, angle=0]{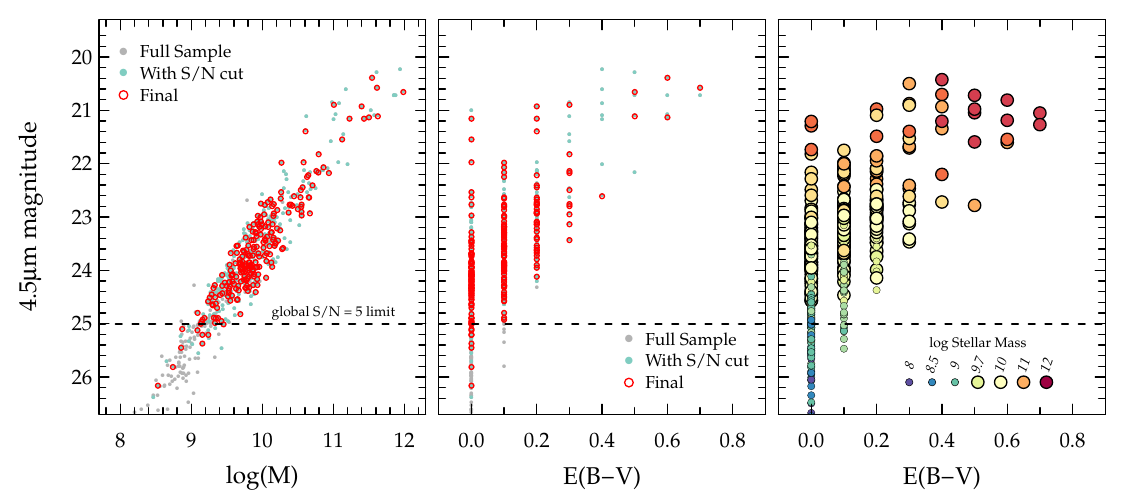}
\caption{\textit{Top:} S/N ratios for galaxies in our sample at $3.9 < z < 4.9$ as a function of their stellar mass (Section~\ref{sec:stellarmass}). The dashed lines indicate the applied conservative S/N cuts (10 at $3.6\,{\rm \mu m}$ and 5 at $4.5\,{\rm \mu m}$) to provide reliable Spitzer/IRAC photometry for the measurement of \halpha~emission. The full sample is shown in gray, the S/N selected sample in turquoise, and the final sample (including rejection of blended sources, Section ~\ref{sec:spitzercontamination}) in red (left and middle panel). The right panel shows both S/N ratios color-coded by stellar mass. Above $\logm\sim9.7$, our sample is complete in terms of stellar mass and the \halpha~measurement is not affected by the imposed S/N selection.
\textit{Bottom:} Similar to the top panels, here showing the $4.5\,{\rm \mu m}$ magnitude (not affected by \halpha) as a function of stellar mass and dust attenuation. \label{fig:sample0}}
\end{figure*}
%%%%%%%%%%%%%%%%%%%%%%%%%%%%%

\subsection{Properties from SED fitting}\label{sec:stellarmass}
For our final set of galaxies, we derive stellar population properties by performing spectral energy distribution (SED) fitting to the $3\arcsec$ aperture photometry provided by the \textit{COSMOS2015} catalog. The photometry is converted to total fluxes and corrected for Galactic foreground extinction \citep{SCHLEGEL98} following the recipe given in \citet{LAIGLE16}. Synthetic SED models are then fit to the photometry points using the public code \texttt{EAZY}\footnote{Version April 2015: \url{http://www.astro.yale.edu/eazy/}} \citep{BRAMMER08}. We note that the use of other SED fitting codes \citep[e.g., \texttt{LePhare},][]{ARNOUTS99,ILBERT06} provide very similar results. 
In detail, we use \citet{BRUZUALCHARLOT03} composite stellar population models with three different stellar metallicities ($Z=0.02$, $0.008$, and $0.004$ where $\Zsol = 0.02$), 15 logarithmically spaced age steps between $8.0 < \log(t/{\rm yr}) < 9.2$, and stellar dust attenuation values between $0 < \ebmvs < 0.7$ in steps of $0.1$ (assuming a Calzetti dust attenuation law). We assume five different SFHs parameterized as constant, exponentially declining, and delayed ($\propto \tau^{-2}\,t\,e^{-t/\tau}$). For the latter two, we assume a timescale $\log(\tau/{\rm yr}) = 8$ and $9$.
To these models, we add \halpha~emission in a grid of different \halpha~EWs from rest-frame $0\,{\rm \AA}$ to $1500\,{\rm \AA}$ in steps of $200\,{\rm \AA}$. Other optical lines are added as described in Section~\ref{sec:emlinfitassumptionslines}.
We do not add emission lines to SED templates for which ${\rm sSFR} < \left(T_{\rm univ}^{\rm z=4}\right)^{-1} \sim 0.6\,{\rm Gyr^{-1}}$ as we consider these galaxies to be quiescent \citep[e.g.,][]{FAISST17a}.
In the following analysis, we will only use the stellar masses and dust attenuation measurements from SED fitting. We have verified that both of these quantities are relatively insensitive to the chosen template set and parameterization of the optical emission lines.

Figure~\ref{fig:sample} shows the stellar mass distribution as a function of redshift for the spectroscopic (filled diamonds) and the photometric (open circles) sample. The red arrows and numbers indicate galaxies with $\logm > 10.5$, whose Spitzer photometry has an increased probability to be affected by blending based on our machine learning approach (see Appendix~\ref{sec:appendixA}).
The inspection of the images and SEDs of these galaxies suggest indeed contamination of the Spitzer filters and we therefore remove those four galaxies in the following.

\subsection{Stellar Mass Completeness}\label{sec:completeness}

The applied S/N cut to obtain reliable \halpha~measurements limits our sample to the brightest, hence most massive galaxies. In the following, we estimate the corresponding mass completeness limit.

Figure~\ref{fig:sample0} summarizes the properties of our sample. The top panels show the S/N ratio at $3.6\,{\rm \mu m}$ and $4.5\,{\rm \mu m}$ as a function of stellar mass. The bottom panels show the $4.5\,{\rm \mu m}$ magnitude (not affected by \halpha) as a function of stellar mass and dust attenuation, respectively. The color-coding in the middle and left panels indicate the full sample (gray, Section~\ref{sec:mainsampleselection}), the sample with applied S/N cuts (turquoise, Section~\ref{sec:snselection}), and the final sample after removing galaxies with contaminated Spitzer photometry (red, Section~\ref{sec:spitzercontamination}). The color-coding on the right panels indicates different stellar masses.

From the top panels, we estimate the stellar mass completeness limit of $\logm \sim 9.7$ (light green dots with border). Above this limit, we do not expect biases cause by our S/N selection.
The bottom panels show a correlation between dust attenuation and stellar mass, as generally expected \citep[e.g.,][]{WHITAKER12}. This cannot be attributed to a selection effect as our sample would include dusty galaxies that are $\sim3$ magnitude fainter. Hence, we expect the most dusty galaxies to be bright, while fainter galaxies commonly exhibit dust attenuation of \ebmv$<0.2$.

%%%%%%%%%%%%%%%%%%%%%%%%%%%%%
%%%		FITTING 		  %%%
%%%%%%%%%%%%%%%%%%%%%%%%%%%%%
\section{Measurement of \halpha~emission} \label{sec:emlinfit}

In this section, we detail our method to measure \halpha~emission using Spitzer broad-bands.
After explaining our approach in detail (Section~\ref{sec:emlinfitmethod}), we outline the assumptions (Section~\ref{sec:emlinfitassumptions}), assess the reliability using mock galaxies (Section~\ref{sec:emlinfittest}), and apply it to our galaxy sample (Section~\ref{sec:emlinfitapplication}).

%%%%%% FIGURE: FIT EXAMPLES %%%%%%%
\begin{figure}
\includegraphics[width=1.0\columnwidth, angle=0]{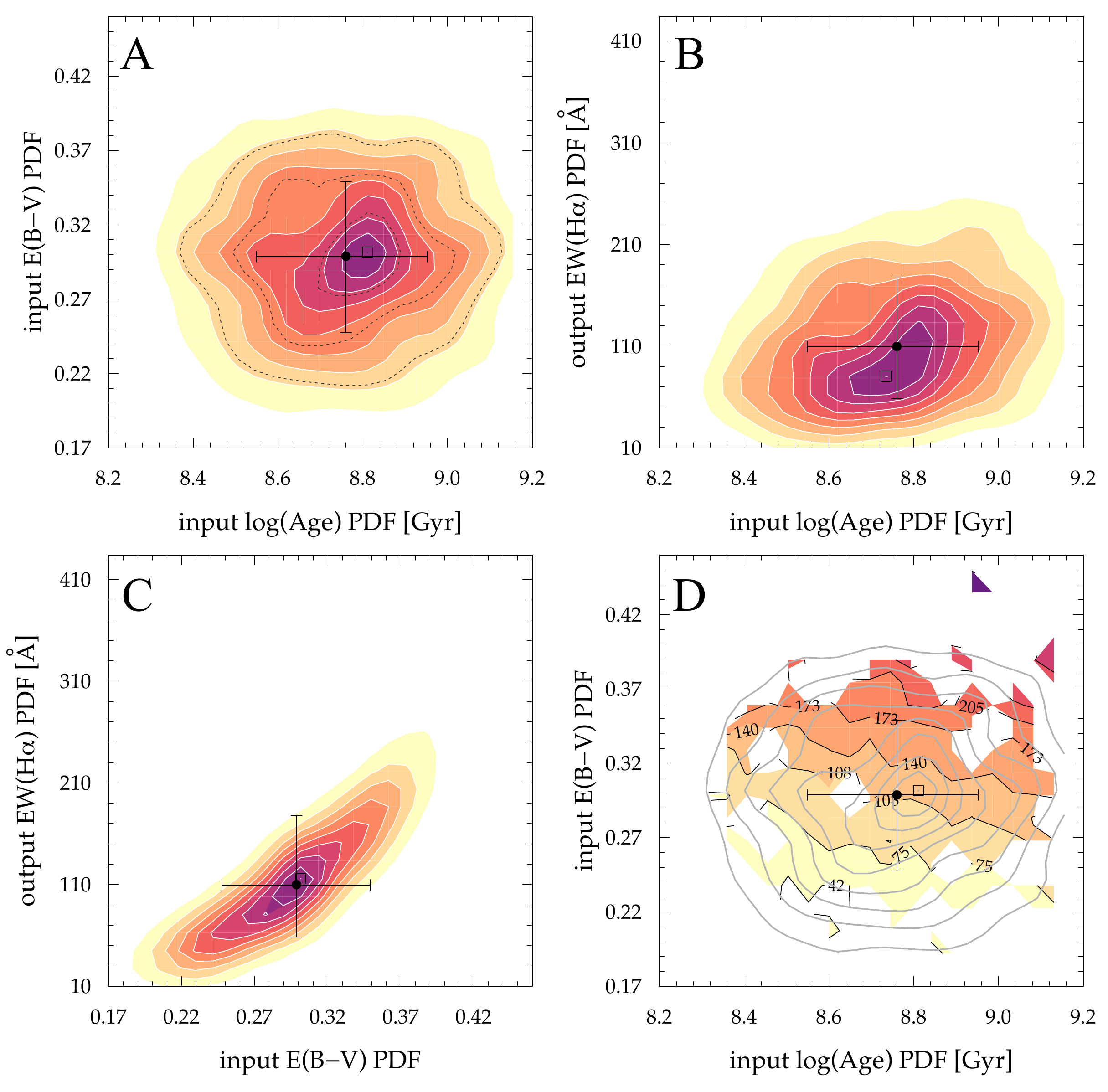}
\caption{Example of a measured of \halpha~luminosities (or EWs) from the \iracA~Spitzer broad-band color for a galaxy at $z=4.6$. In all panels, the square symbol represents the peak of a given PDF, while the point symbol with error bars represents the median of the PDF including $1\sigma$ percentiles. The 2-dimensional input (prior) PDF from dust attenuation and age is shown in Panel A. Panels B and C show the output PDF of the \halpha~EW marginalized over dust attenuation and stellar population age, respectively. Panel D shows the output \halpha~EW PDF contours (color and labeled contours) on the full phase space together with the input PDFs of dust and age (gray contours). The most likely \halpha~EW for this example galaxy is $\sim110\,{\rm \AA}$ with about a $50\%$ $1\sigma$ uncertainty.
\label{fig:fitexample}}
\end{figure}
%%%%%%%%%%%%%%%%%%%%%%%%%%%%%

\subsection{Method}\label{sec:emlinfitmethod}

Optical emission lines in $z>3$ galaxies are out of reach of current spectrographs. However, their measurement using broad-band colors has been well established in the recent past \citep{SHIM11,STARK13,MARMOLQUERALTO16,SMIT15,FAISST16a,RASAPPU16,SMIT16}.
The \iracA~color measured by Spitzer can be used to measure \halpha~emission at $z=4-5$ (Figure~\ref{fig:lines}) at a statistical precision similar to spectroscopy \citep{FAISST16a}. This is possible because the optical continuum in young galaxies ($<1\,{\rm Gyrs}$) at high redshifts (limited by the age of the universe) is relatively insensitive to various stellar population properties.
Our approach builds on the method by \citet{FAISST16a}, which we extend here to take properly into account the uncertainties in observables (color) and model parameters (age, SFH, metallicity, reddening) in a probabilistic way.
Compared to SED fitting of multiple photometric points over a large range of wavelengths, our approach, focusing only on the optical continuum, is faster in obtaining a full realistic probability distribution function (PDF) of the \halpha~emission line flux dependent on the input uncertainties. Our method is advantageous to be applied to large samples of galaxies with well characterized uncertainties.

In detail, we sample from input PDFs including \textit{(i)} the uncertainty in the observed color (here the Spitzer \iracA~color), \textit{(ii)} the age of the stellar populations, and \textit{(iii)} the dust attenuation of the stellar continuum. As shown in \cite{FAISST16a}, the latter two dominantly define the rest-frame optical continuum in galaxies at $z>4$ where the universe is $\lesssim1.5\,{\rm Gyrs}$ old. Other quantities (such as redshift, metallicity, SFH) have a negligible effect on the optical continuum and are varied in discrete steps to investigate their impact on the final results.

The posterior PDF of the \halpha~emission is derived by comparing the observed \iracA~color distribution to model colors measured from several $1000$ optical continua (including emission lines) generated from the prior distribution in age and dust attenuation and a basis set of synthetic templates with varying SFH, stellar metallicity and dust reddening law.
The basis templates are explained in Section~\ref{sec:emlinfitassumptionscontinuum}. The input PDFs for age and dust attenuation are, in our case, obtained from SED fitting (Section~\ref{sec:stellarmass}), but we note that they can also be obtained from other independent sources. The assumptions for emission lines and dust attenuation are outlined in Sections~\ref{sec:emlinfitassumptionslines} and~\ref{sec:emlinfitassumptionsdust}.

Figure~\ref{fig:fitexample} shows an example output of our routine for a galaxy at $z=4.6$ with constant SFH and solar metallicity. A 2-dimensional input PDF for dust attenuation and age is used as shown in Panel A.
The following panels show the output PDF of the \halpha~EW marginalized over dust attenuation (Panel B) and stellar population age (Panel C). Panel D shows the output \halpha~EW PDF contours (color and labeled contours) on the full phase space together with the input PDFs of dust and age (gray contours).
Note the strong dependence of the results on dust attenuation but weak dependence on age.

%%%%%%% TABLE: MODELS  %%%%
\begin{deluxetable}{l c  c c}
\tabletypesize{\scriptsize}
%\rotate
\tablecaption{List of basis composite stellar population models to model the optical continuum.\label{tab:fitinput}}
\tablewidth{0pt}
\tablehead{
\colhead{Model} & \colhead{SFH} & \colhead{Metallicity} & \colhead{Dust attenuation}\\[-0.2cm]
\colhead{} & \colhead{} & \colhead{($Z_{\odot} = 0.02$)} & \colhead{}
}
\startdata
A & constant & 0.02 & Calzetti\\
B & constant & 0.004 & Calzetti\\
C & exp. declining$^{a}$ & 0.01 & Calzetti\\
D & constant & 0.02 & SMC\\
E & constant & 0.004 & SMC\\
F & exp. declining$^{a}$ & 0.01 & SMC\\
% & ($\tau=3\cdot10^8\,{\rm yrs})$ &  & \\
\enddata
\tablenotetext{a}{Assuming $\tau=3\times 10^8\,{\rm yrs}$.}
\end{deluxetable}
%%%%%%%%%%%%%%%%%

\subsection{Assumptions}\label{sec:emlinfitassumptions}

\subsubsection{Optical continuum}\label{sec:emlinfitassumptionscontinuum}
For modeling the optical continuum, we use composite stellar population synthesis models based on the template library by \citet{BRUZUALCHARLOT03} and assume a \citet{CHABRIER03} initial mass function. In order to investigate the effect of different assumptions on our results, we choose $3$ different stellar population models, each with two different parameterizations of dust attenuation (see below). The models include a constant SFH for stellar metallicities of $Z=0.02$ and $Z=0.004$ (where $\Zsol=0.02$) and an exponentially declining SFH proportional to $e^{-t/\tau}$ with $\tau = 3\times10^8\,{\rm yrs}$ and stellar metallicity of $Z=0.01$ (see Table~\ref{tab:fitinput}).
Note that, in contrast to SED fitting, we only need to model the \textit{optical} continuum, which is to first order a well defined power-law and insensitive to most galaxy properties.

%%%%%% FIGURE: EW (2) %%%%%%%
\begin{figure*}[t!]
\includegraphics[width=1.04\columnwidth, angle=0]{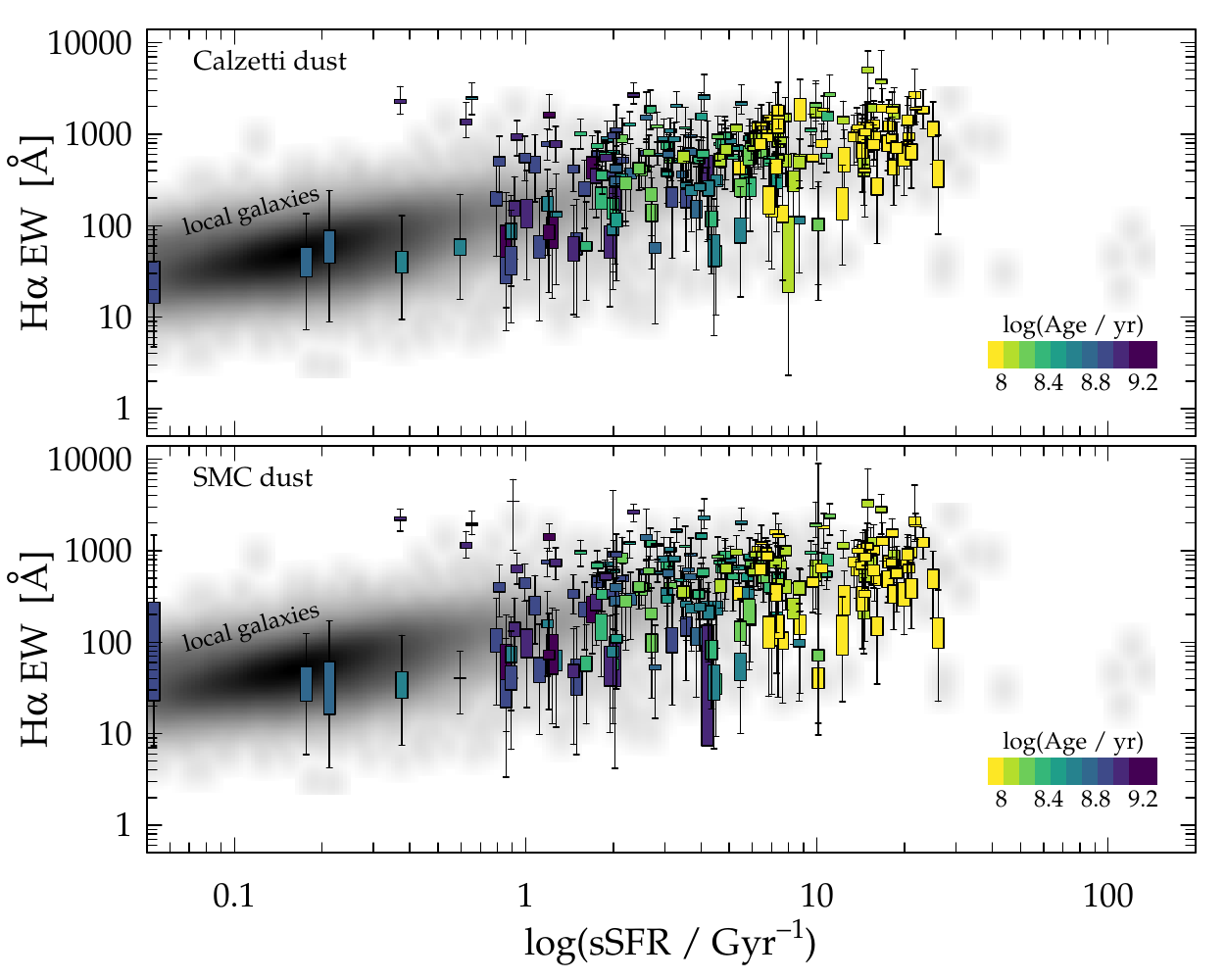}
\includegraphics[width=1.04\columnwidth, angle=0]{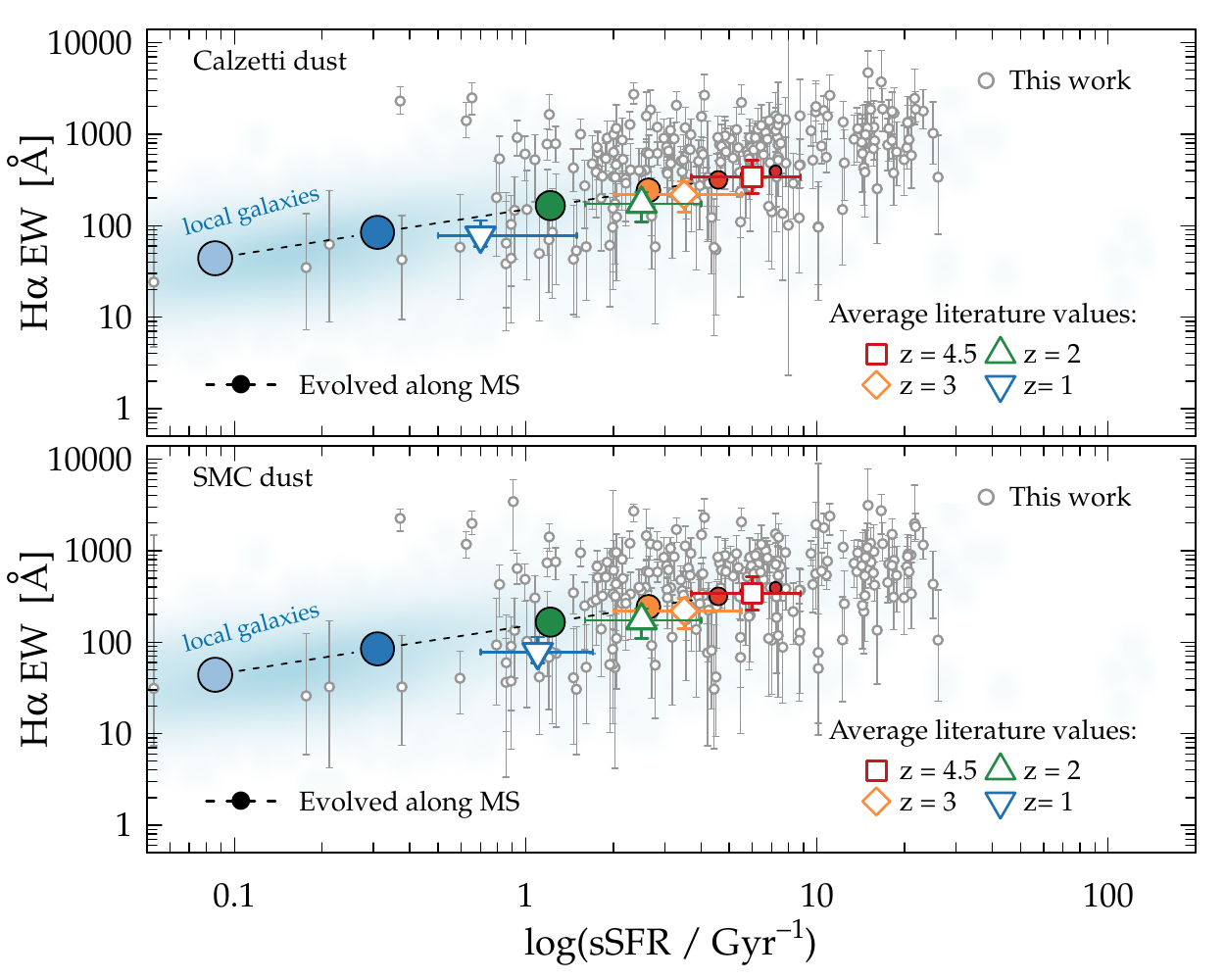}
\caption{Relation between \halpha~EW and sSFR (measured from SED fitting) for Calzetti (top panels) and SMC dust attenuation (lower panels).
\textit{Left:} Our measurements at $z\sim4.5$ (colored by age from SED fitting) compared to local SDSS galaxies (gray cloud). The vertical size of the boxes capture the most likely value for all the models that are used to derive \halpha~emission (see Table~\ref{tab:fitinput}). The error bars indicate the total $1\sigma$ percentiles due to photometric noise.
\textit{Right:} Our measurements (gray points) compared to average measurement from the literature at different redshifts (see text). The dashed line shows path of a model galaxy at $z\sim4.5$ and $\logm=10.0$ evolving along the main-sequence to $z=0$ (same color-code for redshift). Its sSFR decreases over time as its stellar mass is increasing (indicated by size of symbols). The ``redshift invariant'' sequence between \halpha~EW and sSFR hence is expected if the galaxies are evolving along the main-sequence.  \label{fig:ewlocal}}
\end{figure*}
%%%%%%%%%%%%%%%%%%%%%%%%%%%%%

\subsubsection{Emission lines in addition to \halpha}\label{sec:emlinfitassumptionslines}
In addition to \halpha~emission, we add other emission lines in the optical from rest-frame $4000-9000\,{\rm \AA}$. These include \nii~and \siii~as well as weaker lines, which, could affect the observed color. In all cases, the line strengths are coupled to \halpha~in the following ways. For the \niiha~ratio we assume $0.15$, the average for a high-redshift galaxy at a stellar mass of $\logm = 10$ \citep{FAISST18}. For the other lines, we assume relations to \hbeta~(derived from \halpha~flux according to case B recombination) as published in \citet{ANDERS03} for half-solar stellar metallicity. Our results are insensitive to conversions using a different stellar metallicity.

\subsubsection{Dust attenuation and $f$-factor (differential dust reddening between nebular lines and stellar continuum)} \label{sec:emlinfitassumptionsdust}
We assume the standard \citet{CALZETTI00} dust attenuation based on local starbursts without the ``Milky Way bump'' as fiducial dust attenuation model. However, we also explore an alternative extinction law as measured in the Small Magellanic cloud \citep[SMC,][]{PREVOT84,PETTINI98}. The latter may be more suitable for metal-poor galaxies at high redshifts, although the opinion in the literature diverges on the details \citep[e.g.,][]{FUDAMOTO17}.

Furthermore, we assume a fiducial ratio between stellar and nebular attenuation ($f$) of $0.44$ as found in local starburst galaxies \citep{CALZETTI00}\footnote{The $f$-factor is defined as the ratio of stellar continuum to nebular reddening, $f = \frac{\ebmvs}{\ebmvn}$.}. Note that this ratio might be different at high redshifts due to the changing star formation activity and the properties of the ISM of galaxies \citep[e.g.,][]{REDDY15}. The literature has not yet arrived at a consensus on this ratio at higher redshifts, although there are indications of an evolution towards unity at $z\sim2$ \citep{ERB06,HAINLINE09,FORSTERSCHREIBER09,CRESCI12,YABE12,KASHINO13,CULLEN14,PRICE14,THEIOS18}. Local analogs of high-$z$ galaxies further suggest such an evolution (Section~\ref{sec:ffactor}).
In Appendix~\ref{sec:appendixffactor}, we outline how to correct different quantities that are linearly dependent on \halpha~luminosity for varying $f$-factors.

%\subsection{Model biases and dependence on S/N ratio} 
\subsection{Biases from model and observations}
\label{sec:emlinfittest}

In Appendix~\ref{sec:biases}, we provide a comprehensive and detailed study of biases including \textit{(i)} systematic model-based biases of our method in the case of ideal observations (i.e., no photometric noise) and \textit{(ii)} observation-based biases due to the flux limits of the Spitzer surveys.
In this section, we provide a brief summary.

\subsubsection{Biases due to model assumptions}\label{sec:modeldependencetest}

We use a set of simulated SEDs (see Appendix~\ref{sec:biases}) to investigate potential biases introduced by our assumption of the underlying stellar continuum model. Specifically, we compare the true and measured \halpha~EW (similarly \halpha~luminosity) in three cases, namely varying metallicity, SFH, and parameterization of the dust attenuation. These test are done as a function of dust attenuation and age of the stellar population.

We find that assumptions in models that diverge from the true parameterization of the optical continuum SED affect the \halpha~measurements only mildly (less than $20\%$ in most cases). This is a consequence of the weakly varying optical continuum with age and SFH for galaxies at high redshifts where their age is limited to $T_{\rm univ}\sim 1~{\rm Gyr}$. An exception are very dusty galaxies ($\ebmvs > 0.3$) for which these biases are magnified. Specifically, the assumption of an incorrect parameterization of the dust attenuation (SMC vs. Calzetti) can result in errors of the \halpha~measurements of $>50\%$. 
In the following, we derive the \halpha~emission properties for a set of different diverging assumptions (Table~\ref{tab:fitinput}) to reflect these biases in the uncertainty of the final results.

\subsubsection{Biases due to sensitivity limits} \label{sec:sntest}

The sensitivity limits of the Spitzer/IRAC data can affect the \halpha~measurements.
We test this by using the same set of simulated galaxies as in Section~\ref{sec:modeldependencetest}, but add realistic noise to the extracted model photometry (see Appendix~\ref{sec:biases}). We restrict ourselves to an optical continuum modeled by a constant SFH, solar metallicity, and an age of $300\,{\rm Myrs}$. We fix these parameters during fitting in order to be able to compare the effects of photometric noise on our the \halpha~measurements alone. 

We do not find any significant biases in the \halpha~measurements above our imposed S/N limits for dust-poor galaxies ($\ebmvs < 0.2$). However, we expect to significantly overestimate the \halpha~emission by factors $>2$ at \halpha~EWs less than $600\,{\rm \AA}$ for more dusty galaxies with $4.5\,{\rm \mu m}$ magnitudes fainter than $24\,{\rm AB}$ (corresponding to S/N$\sim12$).
Fortunately, dusty galaxies are bright and well detected at $4.5\,{\rm \mu m}$ at S/N$>12$ as shown in Figure~\ref{fig:sample0}.
We therefore conclude that even for very dusty galaxies there is no significant bias of more than a factor of two in the measurement of \halpha~emission properties.

%%%%%% FIGURE: EW (1) %%%%%%%
\begin{figure*}
\includegraphics[width=1.0\columnwidth, angle=0]{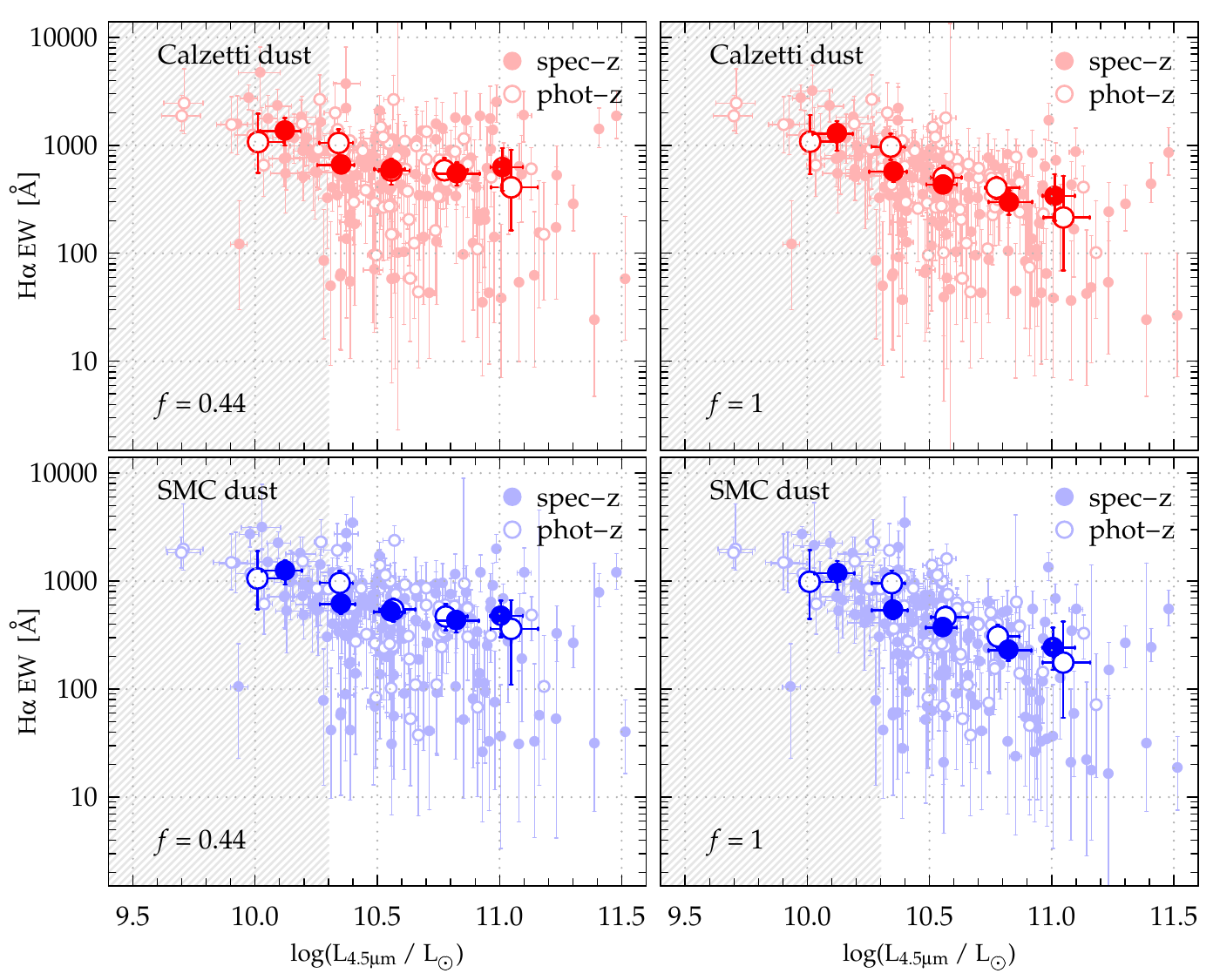}
\includegraphics[width=1.0\columnwidth, angle=0]{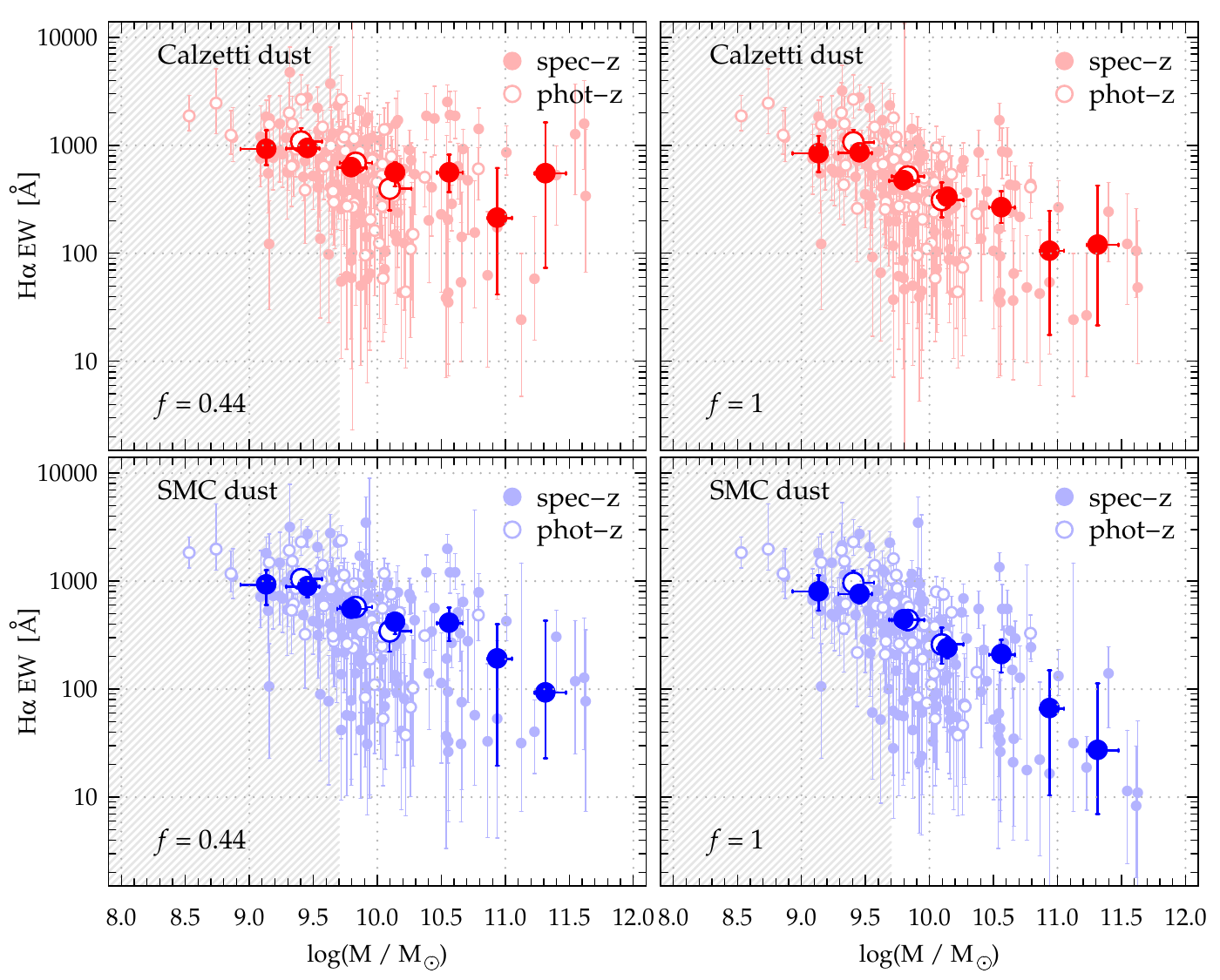}
\caption{\textit{Left:} Relation between \halpha~EW and $4.5\,{\rm \mu m}$ luminosity for our spectroscopic (filled) and photometric (open) sample at $3.9 < z < 4.9$. The four panels show variations in the dust attenuation law and $f$-factor. The large symbols show the weighted means. The gray hatched regions indicates where our sample becomes incomplete, hence misses galaxies with low \halpha~EW. We note an increase in the \textit{maximal} \halpha~EW at the faintest luminosities. On the other hand, our data suggest a suppression of large \halpha~EWs for the brightest luminosities (this trend is enhanced in the case of no differential dust reddening, $f=1$).
\textit{Right:} Same left panel but with stellar mass on the abscissa. Here, the trends are more pronounced. Specifically, we see a suppression of large \halpha~EWs at the highest stellar masses. This could indicate a \textit{temporarily} slow-down of stellar mass build-up in these galaxies.  \label{fig:ew45lummass}}
\end{figure*}
%%%%%%%%%%%%%%%%%%%%%%%%%%%%%

\subsection{\halpha~emission of $z\sim4.5$ galaxies}\label{sec:emlinfitapplication}

To measure the \halpha~properties of our $z\sim4.5$ galaxies, we assume Gaussian input PDFs centered on the best-fit \texttt{EAZY} measurements for the stellar population age and dust attenuation (Section~\ref{sec:stellarmass}), with a $1\sigma$ width of $0.3\,{\rm dex}$ (age) and $0.05\,{\rm mag}$ (in $\ebmvs$), respectively. These values are a fair representation of the accuracy with which these quantities can be estimated with current methods. As described above, we apply two different parameterization for the dust attenuation (local starbursts, and SMC) and assume different stellar population models (Table~\ref{tab:fitinput}) to reflect the model uncertainties in the results. We assume \niiha~$=0.15$ and $f=0.44$ as our fiducial values but we will later extensively discuss the impact of different assumptions.

Figure~\ref{fig:ewlocal} compares the relation between \halpha~EW (from Spitzer photometry) and sSFR (from SED fitting) at different redshifts (for Calzetti and SMC dust attenuation, respectively).

The left panel shows our $z\sim4.5$ measurements colored by age. The height of the boxes include the possible values obtained by the different stellar population models (Table~\ref{tab:fitinput}) and the error bars show the measurement uncertainties (dust, age, and color).
The cloud shows measurements of local galaxies from the Sloan Digital Sky Survey \citep[SDSS,][]{YORK00}. A total sample of $\sim200\,000$ galaxies was extracted using the web-based DR12 \citep{ALAM15} query tool\footnote{\url{ http://skyserver.sdss.org/dr12/en/tools/search/sql.aspx}}. The specific SFRs and \halpha~emission line measurements are taken from the \textit{Galspec} products provided in the MPA-JHU value added catalog and calculated by the methods of \citet{KAUFFMANN03}, \citet{BRINCHMANN04}, and \citet{TREMONTI04}. The \halpha~EWs are dereddened using the Balmer decrement.

The right panel shows these measurements together with the medians of other measurements from the literature color-coded by redshift at $z\sim4.5$ \citep{STARK13,FAISST16a,RASAPPU16,SMIT16,MARMOLQUERALTO16}, $z\sim3$ \citep{MAGDIS10,TASCA15,REDDY12,MARMOLQUERALTO16}, $z\sim2$ \citep{DADDI07a,KARIM11,SOBRAL14,SILVERMAN15,MARMOLQUERALTO16}, and $z\sim1$ \citep{NOESKE07,KARIM11,FUMAGALLI12,SOBRAL14,TASCA15}.
While local galaxies are dominantly represented at low sSFR ($<1\,{\rm Gyr^{-1}}$) and between $10\,{\rm \AA} < {\rm EW}({\rm H\alpha}) < 100\,{\rm \AA}$, the $z\sim4.5$ galaxies occupy large values in both quantities. This is expected from a simple evolution of their star-formation with time: the dotted line shows a $z\sim4.5$ galaxy at $\logm = 10.0$ modeled by \citet{BRUZUALCHARLOT03} composite templates (see Section~\ref{sec:discussion}), which evolves along the star-forming main-sequence \citep{SPEAGLE14}. The large symbols show the galaxy at different redshift (same color code). Note that the sSFR is inconsistent with the measurement from the literature at a given redshift. This is because the model galaxy increases its stellar mass (indicated by the size of the symbols) and hence decreases its sSFR, as it evolves along the main-sequence. The samples from the literature are chosen at a similar stellar mass across all redshifts and therefore show a less strong evolution in sSFR.

It is worth noting that there is a region of overlap between the sSFR and \halpha~EW distribution of local galaxies and galaxies at $z\sim4.5$. This suggests that the ISM conditions present on average in galaxies at high redshifts are similarly present in a subset of galaxies in the local universe. This motivates the use of local galaxies as analogs of high redshift galaxies to study their basic physical properties at early cosmic times \citep[e.g.,][]{FAISST16c}.

%%%%%%%%%%%%%%%%%%%%%%%%%%%%%
%%%		RESULTS 		  %%%
%%%%%%%%%%%%%%%%%%%%%%%%%%%%%
\section{Results} \label{sec:results}

In the following, we detail the main results which include the relation between \halpha~and stellar mass (Section~\ref{sec:ew}), the comparison of \halpha~and UV luminosity (Section~\ref{sec:hauvlum}), the comparison of \halpha~and UV derived SFRs (Section~\ref{sec:hauvsfr}), and the \halpha~main-sequence at $z=4.5$ (Section~\ref{sec:ms}). We also show the effect of model assumptions.

\subsection{EW(\halpha) versus stellar mass}\label{sec:ew}

On the left four panels in Figure~\ref{fig:ew45lummass}, we show the relation between \halpha~EW and the observed $4.5\,{\rm \mu m}$ luminosity. The latter (rest-frame $8000\,{\rm \AA}$) is not affected by \halpha, hence a good proxy for the stellar mass (see Figure~\ref{fig:sample0}).
The different panels show the measurements for a Calzetti (red) and SMC (blue) dust parameterization as well as different $f$-factors ($0.44$ and $1.0$ to capture a reasonable range\footnote{No differential dust attenuation occurs if $f=1$, hence the EW measurement is dust-independent in this particular case.}).
The large symbols show the weighted mean in bins of $L_{4.5{\rm \mu m}}$ with the corresponding asymmetric errors estimated from bootstrapping.
The right four panels of Figure~\ref{fig:ew45lummass} compare the \halpha~EW with stellar mass.

Taken at face-value, the data follow a trend of increasing \halpha~EWs as stellar mass (or equivalently $4.5\,{\rm \mu m}$ luminosity) decreases.
However, several issues have to be pointed out.
First, our sample becomes incomplete at $\log L_{4.5{\rm \mu m}} \sim 10.3$ (10.0) at $z\sim5$ ($z\sim4$) and likewise at $\logm \sim 9.7$ due to the cut in S/N in the Spitzer bands (see Section~\ref{sec:completeness}). This results in a loss of galaxies with low \halpha~EW below these limits. We do not expect biases in the measurements of \halpha~EWs for galaxies above the S/N threshold at these faint luminosities and low stellar masses (see Section~\ref{sec:completeness}). Therefore, that the higher maxima of the \halpha~EWs at low stellar masses are not due to selection and measurement biases.
Second, the negative trend of \halpha~EW with $4.5\,{\rm \mu m}$ luminosity (or stellar mass) is more pronounced in the case of an SMC dust parameterization or a high $f$-factor. This is because more massive galaxies are more dusty, hence $f$ has a larger impact on the measurements of their \halpha~emission.
Third, we note that the \niiha~ratio is on average increasing with stellar mass \citep[e.g.,][]{FAISST18}, which increases the significance of the anti-correlation between stellar mass and \halpha~EW. However, as shown later, this effect is less than $0.15\,{\rm dex}$ and therefore negligible compared to other uncertainties.

Summarizing, the largest uncertainty in our analysis is the unknown $f$-factor. This especially impacts measurements of massive, dust-rich galaxies. Assuming a redshift dependent $f$ (approaching unity as higher redshifts, see also Section~\ref{sec:ffactor}), we see indications of a negative trend between \halpha~EW and $4.5\,{\rm \mu m}$ luminosity or stellar mass. Maximal \halpha~EWs are reached for low-mass galaxies.

%%%%%% FIGURE: Luminosity comparison (1) %%%%%%%
\begin{figure*}[ht!]
\includegraphics[width=2.1\columnwidth, angle=0]{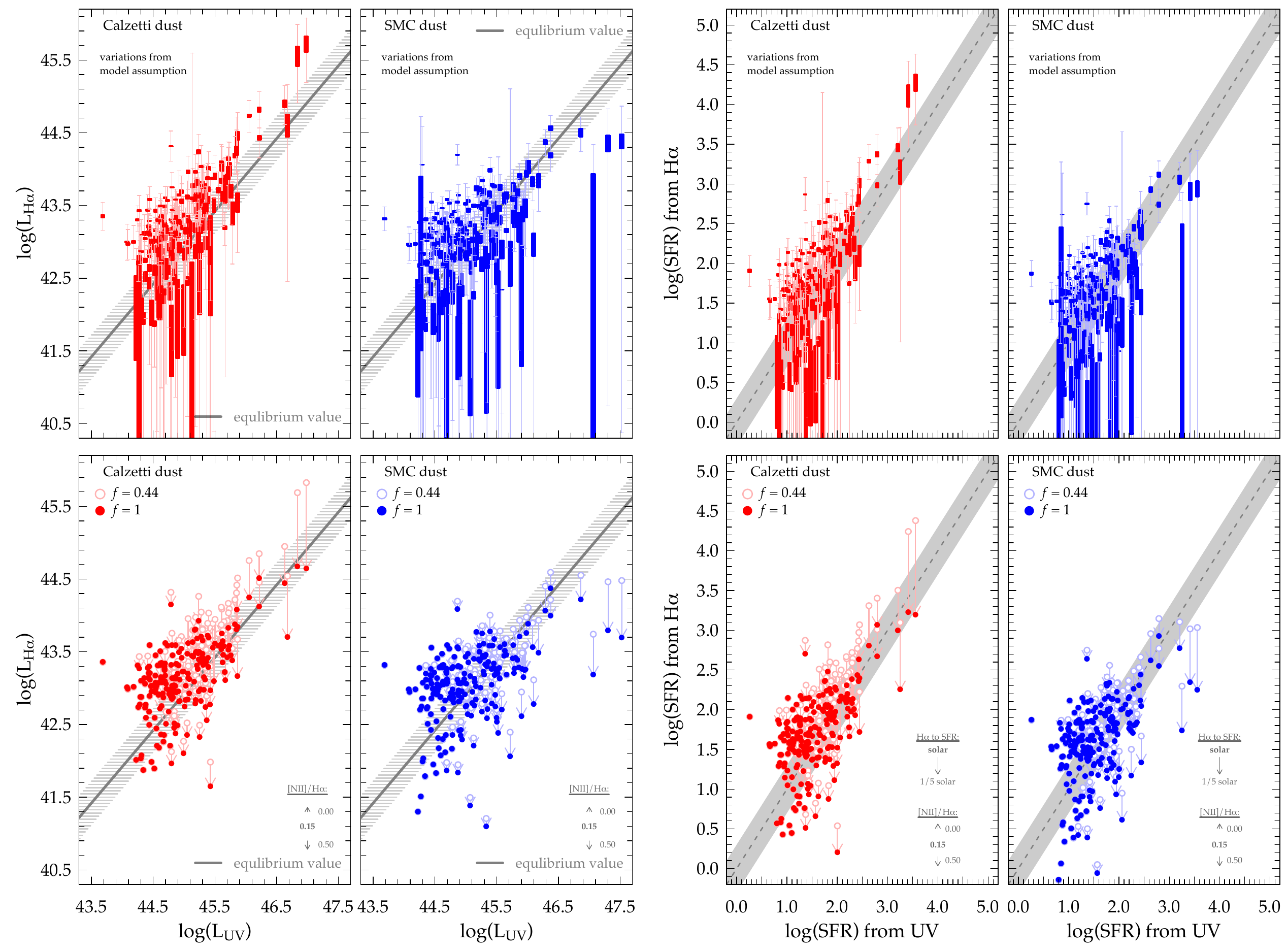}
\caption{\textit{Left four panels:} Comparison of UV luminosity and \halpha~luminosity. The top panels show model (bars) and observational (thin error bars) uncertainties, while the bottom panels show different $f$-factors and the impact of a changing \niiha~ratio. The left and right panels show a Calzetti and a SMC dust attenuation, respectively. The line shows the expected ratio $\LHA/\LUV$ for a constant SFH with a $0.3\,{\rm dex}$ margin. 
\textit{Right four panels:} Comparison of the UV and \halpha~derived SFR with the same layout as the left panel group. In addition, the bottom panels show the impact of different metallicities assumed for the conversion of \halpha~to SFR. The line here shows the ratio SFR$_{\rm H\alpha}$/SFR$_{\rm UV}=1$.
Generally, UV and \halpha~derived quantities are in good agreement, however, we notice a large scatter. Specifically, we find an excess in $\LHA$ and \halpha~SFRs for at least $50\%$ of the galaxies compared to the UV continuum calculated values.
\label{fig:lumsfrcomparison}}
\end{figure*}
%%%%%%%%%%%%%%%%%%%%%%%%%%%%%

%%%%%% FIGURE: Main Sequence %%%%%%%
\begin{figure*}
\includegraphics[width=2.1\columnwidth, angle=0]{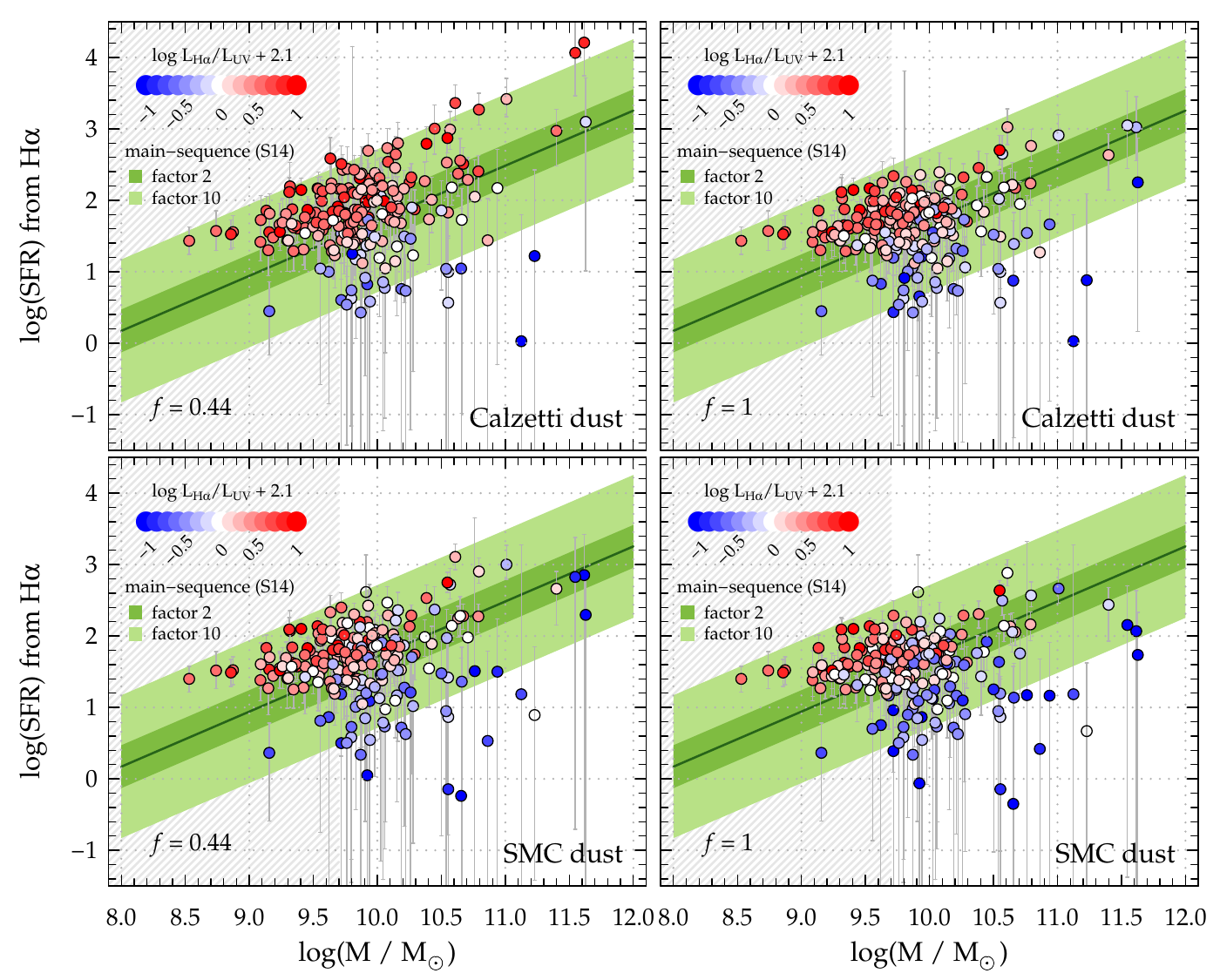}
\caption{Main-sequence of $z\sim4.5$ galaxies (SFR derived from our \halpha~measurements). The error bars take into account photometric errors as well as model uncertainties (c.f. previous figures). The points are color-coded by the ratio of \halpha~to UV luminosity normalized to the ``equilibrium'' value of $-2.1$ (red: \halpha-dominated; blue: UV-dominated; white: consistent with constant star formation). The different panels show different reddening curves and $f$-factors. The green line shows the main-sequence parameterization in \citet{SPEAGLE14} at $z=4.5$, the bands have a width of a factor of $2$ (dark green) and $10$ (light green) respectively. The scatter in the \halpha-derived main-sequence is due to varying ratios of \halpha~to UV SFRs, which can be interpreted as different recent SFHs of the galaxies (see Section~\ref{sec:discussion}).
\label{fig:mainsequence}}
\end{figure*}
%%%%%%%%%%%%%%%%%%%%%%%%%%%%%

%%%%%% FIGURE: Main Sequence %%%%%%%
\begin{figure}
\includegraphics[width=1.0\columnwidth, angle=0]{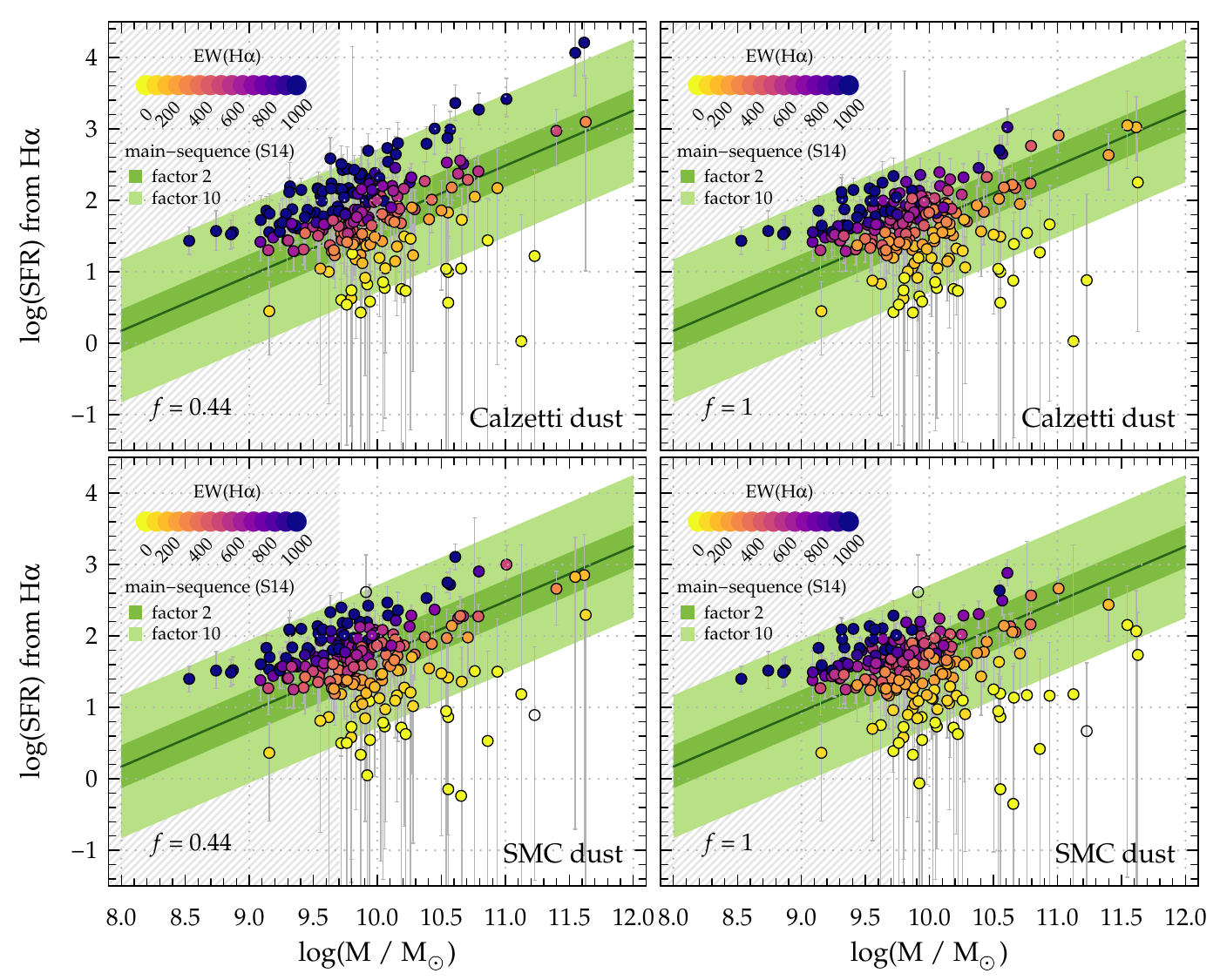}
\caption{Same as Figure~\ref{fig:mainsequence}, but here the points are color-coded by \halpha~EW.
\label{fig:mainsequenceew}}
\end{figure}
%%%%%%%%%%%%%%%%%%%%%%%%%%%%%

\subsection{Comparison of \halpha~and UV luminosity}\label{sec:hauvlum}

The left four panels of Figure~\ref{fig:lumsfrcomparison} compare the \halpha~and UV luminosity\footnote{The UV luminosity ($\LUV$) is defined as $\nu_{1600}{ L}_{1600}$ at rest-frame $1600\,{\rm \AA}$.}. The latter is directly derived from rest-frame $1600\,{\rm \AA}$ emission (which corresponds roughly to the $z$-band at $z\sim4.5$) and subsequently reddening corrected using the stellar $\ebmvs$ values derived from SED fitting (Section~\ref{sec:stellarmass}).
In the top panels (one for each dust parameterization), the height of the boxes represents the uncertainty introduced by the different models and the error bars include the photometric uncertainty. The bottom panels show specifically the uncertainties introduced by different $f$-factors. Higher $f$-factors decrease the \halpha~luminosity of massive, dusty galaxies significantly.
The gray line shows the expected correlation between \halpha~and UV emission in the case of equilibrium, i.e., a constant rate of star formation for several $100\,{\rm Myrs}$, with a margin of $0.3\,{\rm dex}$ (see Section~\ref{sec:discussion}).
Our data suggest that at least $50\%$ of the galaxies have an excess in \halpha~emission with respect to that equilibrium value. For at least $30\%$ of galaxies, this excess is more than a factor of two. Similarly as before, this statement is robust for galaxies at low masses, however, depends on the assumed $f$-factor at higher stellar masses.

\subsection{Comparison of SFRs from \halpha~and UV}\label{sec:hauvsfr}

The comparison of SFRs derived from \halpha~and UV emission directly visualizes the burstiness of star-formation activity. However, we note that the conversion from \halpha~luminosities to SFRs is afflicted with several uncertainties including unknown metallicity, stellar populations, or IMF \citep[][]{KENNICUTT83}. We therefore make the reader aware that the comparison between \halpha~and UV \textit{luminosity} is a more robust measure. Nonetheless, we present the comparison of SFRs here for completeness and at the same time highlight the caveats.

To convert the (reddening-corrected) Spitzer-derived \halpha~luminosities to SFRs, we use the conversion factor originally introduced in \citet{KENNICUTT98}, which is valid for galaxies with solar metallicity.
The UV SFRs are derived from the reddening-corrected UV luminosities (Section~\ref{sec:hauvlum}) using the normalization provided by the same paper.

The right four panels in Figure~\ref{fig:lumsfrcomparison} compare the SFRs measured from \halpha~and UV. Similar to the left four panels, the upper panels focus on the uncertainties from the model, while the lower panels illustrate the effects of various other assumptions. The gray line shows the 1-to-1 relation (with $0.3\,{\rm dex}$ margin) between the SFRs (i.e., the equilibrium value for a constant star formation).
The \halpha~SFRs generally follow well the UV-derived SFRs and the scatter in the relation is similar to what was previously found \citep{SMIT16}. However, it is worth noting that $30-50\%$ of galaxies show a significant excess (factor $>2$) in \halpha-SFRs with respect to the equilibrium, in agreement with the result in Section~\ref{sec:hauvlum}.

Several caveats have to be pointed out. They are quantitatively visualized in the lower right panels of Figure~\ref{fig:lumsfrcomparison} and in more detail described in Section~\ref{sec:discussion}. Most of them do not have a significant impact on the final results.
First, it is expected that the \niiha~line ratio increases with stellar mass \citep[e.g.,][]{FAISST18}. However, even extreme ratios ($0$ or $0.5$) that differ substantially from our fiducial value of $0.15$ result in changes in the SFR of less than $0.1\,{\rm dex}$ in either direction.
Second, at low metallicities, the conversion between \halpha~luminosities and SFRs is suggested to be different from the standard \citet{KENNICUTT98} factor \citep{LY16}. This becomes particularly important for galaxies at high redshifts and low stellar masses (hence low SFR), whose metal content drops below solar \citep[][]{ANDO07,MAIOLINO08,FAISST16b,CULLEN19}. At one-fifth of solar metallicity, the inferred SFRs are expected to be $\sim0.2\,{\rm dex}$ lower \citep{LY16}. Note that this factor is not enough to remedy the large \halpha~luminosity excess of these galaxies.
The largest uncertainty comes from the $f$-factor and dust parameterization. In all cases other than a Calzetti dust parameterization and $f=0.44$, the most star-forming galaxies show a deficit in \halpha~SFR. A good (or any) constraint on $f$ is therefore crucial for the interpretation of these data (Section~\ref{sec:discussion}). JWST will be able to provide constraints on $f$ at $z>4$ by the measurement of Balmer decrements.

Summarizing, we find that at least $50\%$ of galaxies have increased \halpha~SFR with respect to the UV and the scatter is large ($0.5\,{\rm dex}$) around the equilibrium value. This conclusion is robust for low-mass (low SFR) galaxies that are less dusty, hence less affected by uncertainties in the assumptions of the dust parameterization and $f$-factor. Uncertainties in the conversion of \halpha~luminosity to SFR as well as the \niiha~ratio have no significant effect on these results.

\subsection{The \halpha~main-sequence at $z=4.5$}\label{sec:ms}

Figures~\ref{fig:mainsequence} and~\ref{fig:mainsequenceew} show our galaxies on the star-forming main-sequence. The four panels in each figure show different parameterizations of dust attenuation and $f$-factors. The error bars include the photometric measurement error and the uncertainties from the optical continuum models.
In contrast to other measurements of the main-sequence at $z\sim4.5$ that show the SFR derived from rest-frame UV continuum \citep[green line from ][]{SPEAGLE14}, we show the SFR derived from \halpha.
On average, the \halpha~derived main-sequence is in good agreement with the one derived from rest-frame UV continuum at $z=4.5$, however, we notice a large scatter of up to a factor of $10$ (light-green band).
Keeping the caveat of small sample statistics in mind, we note that the scatter is best described with a skewed Gaussian distribution with a tail reaching up to $20$ times below the UV main-sequence for galaxies at $\logm>10.5$.
As shown in Section~\ref{sec:discussion}, such a scatter can naturally be explained by a bursty SFH \citep[see also][]{DOMINGUEZ15}.

The color-coding of the symbols in both figures reflects the results found in the previous sections. Specifically, the colors in Figure~\ref{fig:mainsequence} show changes in the $\LUV/\LHA$ ratio (normalized to an equilibrium value, see Section~\ref{sec:discussion}) across the main-sequence. As expected, galaxies on the main-sequence show comparable UV and \halpha~SFRs (equilibrium value).
The colors in Figure~\ref{fig:mainsequenceew} represent the range of \halpha~EWs. Galaxies on the main-sequence have on average EWs of $\sim200-500\,{\rm \AA}$, in agreement with models in Figure~\ref{fig:ewlocal}. Galaxies at high stellar masses have lower EWs and tend to fall below the main-sequence (assuming an evolving $f$-factor).

\section{Discussion}\label{sec:discussion}

Our main results are the following.

\begin{itemize}
\item A tentative negative correlation between \halpha~EW and stellar mass (or $4.5\,{\rm \mu m}$~luminosity). This correlation is more pronounced if a redshift-dependent $f$-factor is assumed (Figures~\ref{fig:ew45lummass} and~\ref{fig:mainsequence}).
\item A higher ratio of \halpha~to UV continuum luminosity w.r.t. the expected value for a constant star formation (``equilibrium condition'') for at least $50\%$ of galaxies. However, the most massive galaxies show less excess in \halpha~or even a deficit if a redshift dependent $f$-factor is assumed (Figures~\ref{fig:lumsfrcomparison} and~\ref{fig:mainsequenceew}).
\end{itemize}

Before interpreting these results in detail in Section~\ref{sec:bursty}, we first discuss in more detail the redshift-dependence of the $f$-factor (Section~\ref{sec:ffactor}) and the impact of assumptions and measurement uncertainties on the significance of an excess (or deficit) in \halpha~compared to UV continuum emission (Section~\ref{sec:isexcessphysical}).

\subsection{What is the most likely $f$-factor at $z\sim4.5$?}\label{sec:ffactor}

The unknown differential dust attenuation between nebular regions and stellar continuum ($f$) is the main uncertainty in the interpretation of our results. Specifically, it directly affects trends as a function of stellar mass as more massive galaxies are more dusty, hence their reddening correction is more dependent on $f$.
Constraining $f$ at these redshifts is therefore key.

The $f$-factor is commonly measured by comparing the reddening derived from continuum (SED fitting or UV slope) and from the Balmer decrement (\halpha/\hbeta~ratio). The latter can currently not be measured at $z>2.5$.
Several studies suggest that the $f$-factor may significantly change depending on the star-formation activity of galaxies (e.g., sSFR or SFR surface density), which alters the fraction of diffuse ISM to short-lived, dusty stellar birth-clouds \citep{PRICE14,REDDY15}.
Since the average sSFR increases with redshift \citep[e.g.,][and references therein]{FAISST16a}, it is reasonable to expect that $f$ changes as a function of cosmic time.
Several observational studies suggest that $f$ deviates from the average local value of $0.44$ and approaches unity\footnote{The current value ranges from $0.7$ to $0.8$} at $z=2$ \citep{ERB06,REDDY10,KASHINO13,KOYAMA15,VALENTINO15,KASHINO17}.
JWST will be crucial in confirming this trend to higher redshifts and to study its dependence on other galaxy properties.

\begin{figure}
\includegraphics[width=1\columnwidth, angle=0]{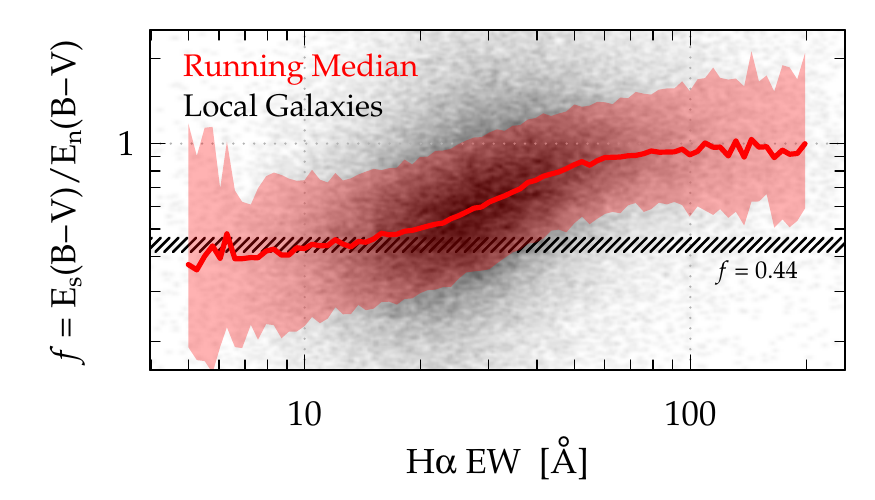}
\caption{The $f$-factor (differential dust reddening between nebular line emission and stellar continuum) as a function of \halpha~EW for low-redshift SDSS galaxies (gray cloud). The running median with $1\sigma$ range is indicated in red. A positive deviation from $f=0.44$ at high \halpha~EWs is indicated by these data. High-redshift galaxies have high \halpha~EWs, which could suggest to similar ISM conditions with high $f$-factors.  \label{fig:ffactor}}
\end{figure}

Assuming that $f$ is coupled to a change in the star formation environment, we would, for example, expect $f$ to change as a function of \halpha~EW (directly correlated to sSFR, Figure~\ref{fig:ewlocal}) also in local galaxies.
To perform a first-order test, we use the sample of local galaxies introduced in Section~\ref{sec:ew}. The dust attenuation of the stellar continuum is measured using the dust opacity in V-band \citep[$\tau_{\rm v}$,][]{BLANTON05}. The nebular dust attenuation is derived from the Balmer decrements.
Figure~\ref{fig:ffactor} shows the $f$-factor as a function of dust-corrected \halpha~EW for this local sample. Qualitatively, we see a positive trend between $f$ and $\ewha$\footnote{A similar trend is seen by relating $f$ to sSFR.}.
As the population-averaged $\ewha$ is increasing on average towards higher redshifts (c.f. Figure~\ref{fig:ewlocal}), this result verifies the observational trends seen out to $z=2$ and suggests their continuation to higher redshifts.
Taking this result at face value, we therefore prefer $f > 0.44$ at $z\sim4.5$ in the following discussion.

%%%%%% FIGURE: F-FACTOR OF LOCAL GALAXIES %%%%%%%
%%%%%%%%%%%%%%%%%%%%%%%%%%%%%

%say what f-factor is the best and if SMC or more Calzetti like.

%%%%%% FIGURE: SFR MODEL %%%%%%%
\begin{figure*}
\includegraphics[width=1\columnwidth, angle=0]{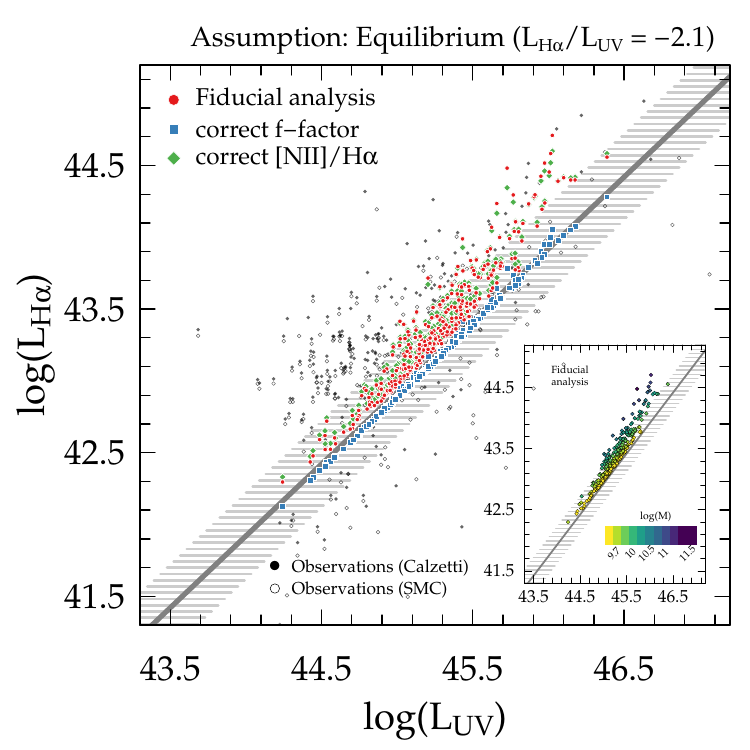}
\includegraphics[width=1\columnwidth, angle=0]{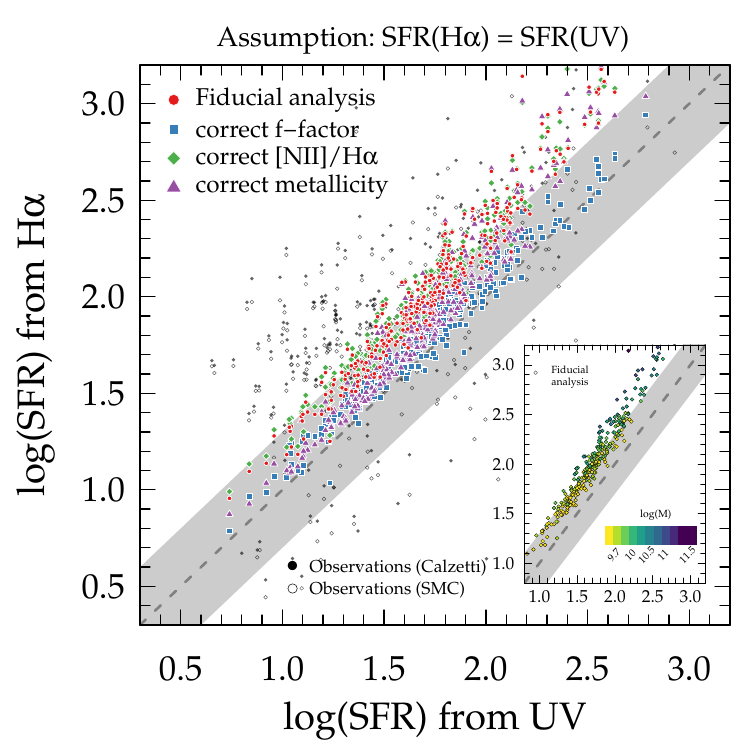}
\caption{Effect of uncertainties in our fiducial assumptions on the derivation and interpretation of \halpha~luminosity (left) and SFRs (right). We assume model galaxies in equilibrium (constant SFR for a long time, equilibrium ratios indicated by diagonal lines) with realistic mass-SFR relation, metallicities, and \niiha~ratios and an $f=0.8$. The red points show the result when analysis these model galaxies using our fiducial set of parameters ($f=0.44$, solar metallicity for the conversion of \halpha~luminosity to SFR, and an \niiha~value of $0.15$). We would overestimate the ``true'' \halpha~luminosities and SFRs substantially ($>0.5\,{\rm dex}$) for the most massive and luminous galaxies due to uncertainties in the $f$-factor. At low stellar masses and luminosities, these uncertainties result in only small biases.
The observations are shown in black (filled: Calzetti; open: SMC). At high stellar masses and luminosities, they coincide with the analyzed model galaxies, which suggests that they could be affected by our assumptions of parameters, hence have a $\LUV/\LHA$ and SFR ratio close to the equilibrium values (or below). On the other hand, the excess in \halpha~luminosity and SFR with respect to the equilibrium ratio in low-mass galaxies cannot be fully explained by deviations from our choice of parameters. \label{fig:sfrsim}}
\end{figure*}
%%%%%%%%%%%%%%%%%%%%%%%%%%%%%

\subsection{Is the excess in \halpha~star-formation physical?}
\label{sec:isexcessphysical}

Next, we discuss in detail the effect of various assumptions made in our model on the relation between \halpha~and UV continuum emission.
Anticipatory, we show in Figure~\ref{fig:sfrsim} that all possible uncertainties (\niiha, metallicity, $f$-factor) cannot remedy the positive excess of \halpha~luminosity (and SFRs) with respect to UV luminosity (and SFRs) of galaxies with $\log({\rm SFR}/ M_{\odot}{\rm yr}^{-1}) < 2$ (corresponding to $\logm \lesssim 11$). This excess has therefore to be physically motivated (Section~\ref{sec:bursty}). At higher stellar masses, however, various uncertainties affect the results.

In the following, we detail our reasoning.
We create a fiducial set of model galaxies for which we assume a ratio of \halpha~and UV continuum emission as expected for a constant SFH without any bursts. We then ``observe'' the model galaxies with the same assumptions of $f$, \niiha~ratio, and metallicity that are applied to the real data. This allows us to verify if the excess in \halpha~is real or caused by our assumptions.
We draw $200$ galaxies from the stellar mass function at $3.5 < z < 4.5$ \citep{DAVIDZON17}, with stellar masses between $9.7 < \logm < 12$. To these we assign (UV continuum) SFRs as a function of their stellar mass using the relation by \citet{SPEAGLE14}. We apply a Gaussian scatter of $0.3\,{\rm dex}$ as measured for the star-forming main-sequence \citep[e.g.,][]{DADDI07a,RODIGHIERO10,WHITAKER12,GUO15,ILBERT15,DAVIES16,KATSIANIS16,SCHREIBER15}.
Using the stellar mass versus gas-phase metallicity relation at $z\sim3.5$ \citep{MAIOLINO08}, we apply metallicities to our galaxies, assuming a Gaussian scatter of $0.2\,{\rm dex}$ to mimic measurement uncertainties and physical scatter. Dust attenuation values in $\ebmvs$ are assigned as a function of stellar mass to match real galaxies at $4 < z < 5$ from the \textit{COSMOS2015} catalog. Finally, we compute \halpha~luminosities from the \halpha~SFRs\footnote{Which are here identical to the UV SFRs by definition.} using the metallicity-dependent normalization by \citet{LY16}. A contribution from \nii~\citep[using the parameterization with stellar mass from][]{FAISST18} is also included to mimic the blending of the lines. Assuming a constant value of $f = 0.8$ \citep[e.g.,][]{KASHINO17}, we redden the \halpha~and UV continuum luminosities to obtain observed quantities.

The red circles in the panels of Figure~\ref{fig:sfrsim} show the relation between \halpha~and UV luminosities and SFRs of the model galaxies, analyzed with the same fiducial parameter as the real galaxies ($f=0.44$, \niiha~=~0.15, and the standard Kennicutt conversion factor between \halpha~luminosity and SFR). Our observed galaxies are shown in black. Because we assumed that the model galaxies are in ``star-formation equilibrium'' (i.e., fixed ratio of $\log \LHA/\LUV \sim -2.1$ and ${\rm SFR_{H\alpha}}/{\rm SFR_{UV}}=1$), deviations from this expectation shows the biases introduced by our (simplistic) choice of fiducial parameters.
From this, two important conclusions can be made.
First, an incorrect $f$-factor (here $0.44$ instead of $0.8$) has a large impact on the \halpha~SFRs of massive ($\logm > 11$, see inset) galaxies. If $f$ is indeed enhanced for $z\sim4.5$ galaxies, our observations at $\logm > 11$ are consistent with no excess in \halpha~(or even a deficit if $f>0.8$).
Second, at lower SFRs (roughly $\log({\rm SFR}) < 2$), the model galaxies observed with our fiducial parameters cannot reproduce the large excess and scatter seen in the observed data (even including the photometric uncertainties). Hence for those galaxies, the excess in \halpha~emission is not caused by inaccurate model assumptions.

For completeness, we also show the \halpha~SFR excess broken up into assumptions other than the $f$-factor.
A correct \niiha~ratio would not improve the observations by much (green diamonds show analysis with correct, mass-dependent ratio).
A metal-dependent conversion of \halpha~luminosity to SFR would improve the measurements at low SFR (galaxies with low-metallicity), however not at high SFRs (purple triangles).

%%%%%% FIGURE: SFH MODELS AND SFR_Ha vs SFR_UV %%%%%%%
\begin{figure*}
\includegraphics[width=2.1\columnwidth, angle=0]{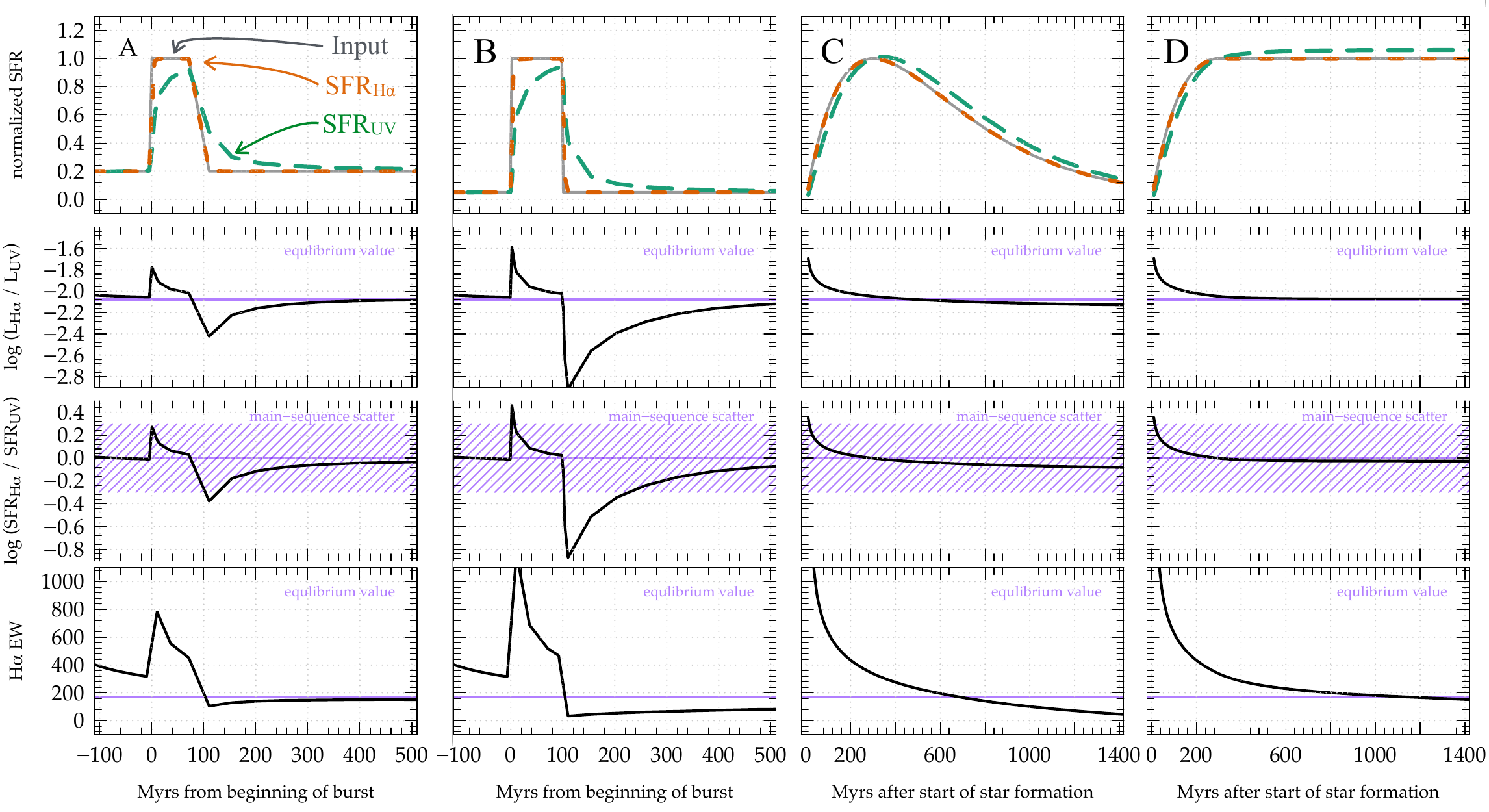}
\caption{Four different SFHs (panels A through D, see text) for which the top sub-panels show the input (gray, solid), \halpha~(orange, short dashed), and UV (green, long dashed) SFRs, respectively. Also shown is the time evolution of the ratio of \halpha~to UV~luminosity (second from top), \halpha~to UV-based SFR (third from bottom) and \halpha~EW (bottom) for each of them. Next to smooth SFHs (C and D), we show two single bursts (A and B) with a duration of $100\,{\rm Myrs}$ and a $5$ and $20$-fold increase in SFR, respectively. All bursts occur after $300\,{\rm Myrs}$ of constant star formation. The different timescales of star-formation probed by \halpha~and UV light results in changes of their ratio as well as the \halpha~EW. In equilibrium (constant SFR), the logarithmic \halpha~and UV-luminosity ratio approaches $-2.1$, the \halpha~and UV SFRs become comparable, and the \halpha~EW approaches $\sim170\,{\rm \AA}$.
\label{fig:bursty}}
\end{figure*}
%%%%%%%%%%%%%%%%%%%%%%%%%%%%%

%%%%%% FIGURE: SFH MODELS ON THE MAIN-SEQUENCE %%%%%%%
\begin{figure*}
\includegraphics[width=2\columnwidth, angle=0]{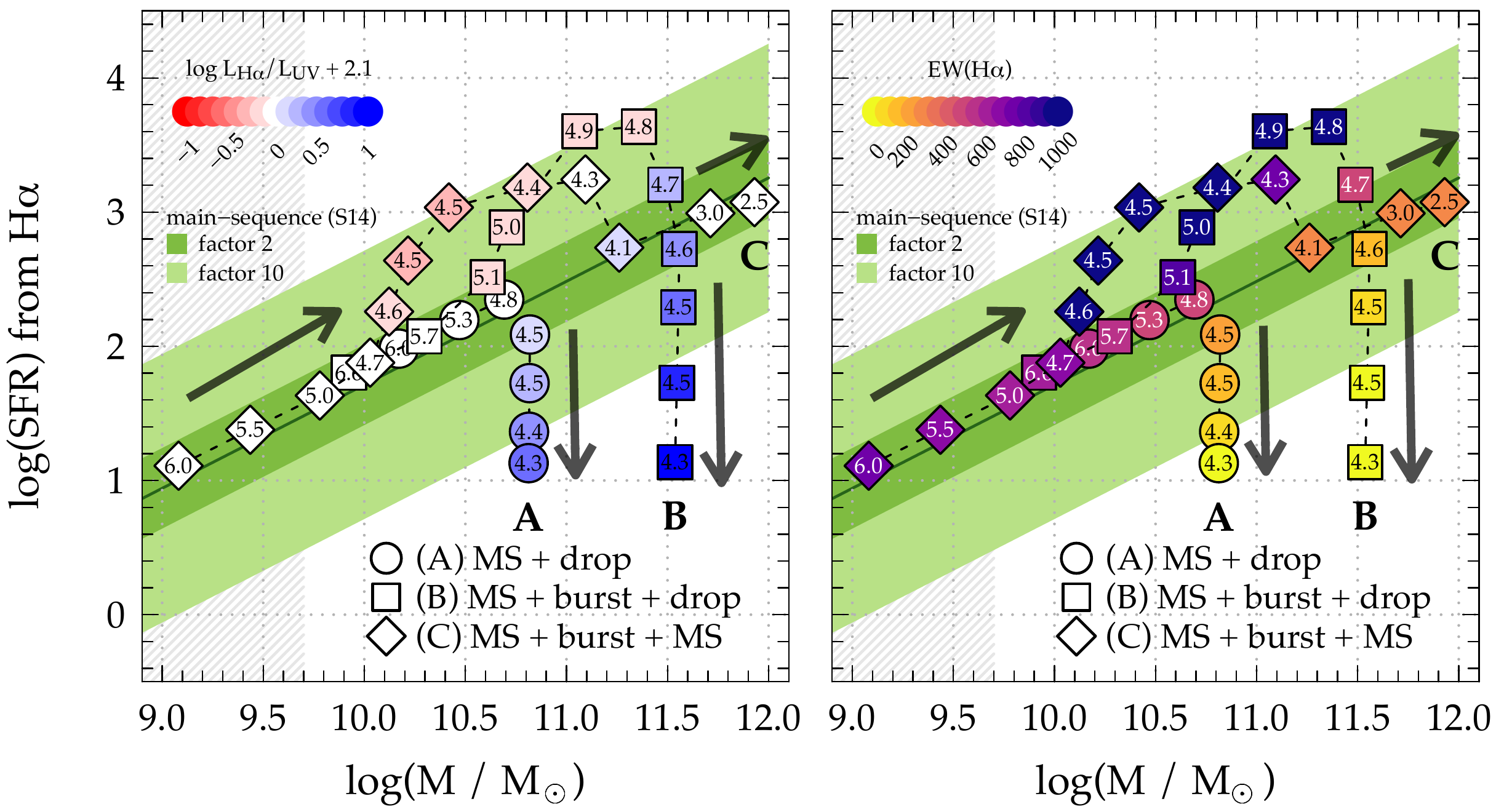}
\caption{Redshift evolution of three model galaxies on the SFR versus stellar mass plane, color-coded by the ratio of \halpha~to UV luminosity (left, normalized to equilibrium value) and \halpha~EW (right). The redshifts are indicated by the small numbers. Model galaxy A grows on the main-sequence and experiences a sudden drop in star-formation at $z=4.5$ to $20$ times below the main-sequence. Model galaxy B grows on the main-sequence but experiences a burst of $20$ times its average SFR at $z=5$ with a duration of $100\,{\rm Myrs}$ before it drops a factor of $20$ below the main-sequence. Galaxy C experiences a burst ($20$ times mean SFR) and continues to grow along the main-sequence.  The galaxies reach a stellar mass of $\logm=10.8$, $11.6$, and $11.4$, respectively, at $z\sim4.1$. Note that the color-coding on the left panel is identical to Figure~\ref{fig:mainsequence} and the green shadings show the main-sequence at $z=4.5$ using the parameterization of \citet{SPEAGLE14} with different margins.
\label{fig:mainsequencemodel}}
\end{figure*}
%%%%%%%%%%%%%%%%%%%%%%%%%%%%%

\subsection{Bursty star-formation in $z\sim4.5$ galaxies}\label{sec:bursty}

Galaxies are expected to oscillate around the main-sequence in a region of roughly $\pm0.3\,{\rm dex}$ as their star-formation activity changes due to gas inflow/outflow and minor/major mergers. The oscillation period has been constrained by cosmological simulations to be $\sim1\,{\rm Gyr}$ at $z=3$, but is likely shorter at higher redshifts and lower masses \citep[e.g.,][]{TACCHELLA16}.
\halpha~emission is tracing the instantaneous star-formation activity and is therefore sensitive to changes in SFR on much shorter timescales and with potential amplitudes much larger than $0.3\,{\rm dex}$. These changes could be caused by feedback mechanisms (supernovae, AGN) acting on the star-formation activity on shorter times \citep[e.g.,][]{HOPKINS14,SPARRE17}.

In Section~\ref{sec:isexcessphysical}, we argue that the excess in \halpha~emission relative to an equilibrium stellar population cannot be caused by model or measurement uncertainties for galaxies up to a stellar mass of $\logm \sim 11$. For more massive and dusty galaxies, the case is less clear due to the unknown $f$-factor. Our data suggest that the majority of massive galaxies in our sample show little \halpha~enhancement or even a deficit if assuming a redshift-dependent $f$-factor (see also Section~\ref{sec:ffactor}).
We now interpret these results, in particular Figures~\ref{fig:mainsequence} and \ref{fig:mainsequenceew}, in the light of the recent star-formation activity of these galaxies.

\subsubsection{Expected correlation between \halpha~and UV continuum from models}
We use simple composite stellar population models to estimate the ratio of \halpha~and UV continuum luminosity as well as \halpha~EW for different SFH scenario. This knowledge is then used to interpret our observations in Section~\ref{sec:interpretationofobservations}.

The simple models discussed here are based off time-evolving composite stellar population models created by the \texttt{GALAXEV}\footnote{\url{http://www.iap.fr/useriap/charlot/}} code. We assume the \citet{BRUZUALCHARLOT03} model library and implement a \citet{CHABRIER03} IMF with a metallicity of $Z=0.008$ (i.e., half-solar). From the model composite stellar population SEDs, the UV SFR is measured from the GALEX FUV photometry ($\sim1500\,{\rm \AA}$) and derived using the conversion by \citet{KENNICUTT98}. The \halpha~luminosity is measured from the rate of H-ionizing photons assuming case B recombination and an intrinsic escape fraction of $100\%$. From this, the \halpha~SFRs are computed using the metal-dependent conversion derived in \citet{LY16}.

For illustration, Figure~\ref{fig:bursty} shows the output for four different SFHs. As a function of time, the normalized SFR, the ratio of \halpha~to UV luminosity and SFR, and the \halpha~EW are shown.
Two of the SFHs (panels A and B) consist of a single bursts, occurring after $300\,{\rm Myrs}$ of constant star formation, with a duration of $100$ Myrs and a $5$ and $20$-fold increase in star-formation, respectively. The two other SFHs (panels C and D) show smoothly rising and falling SFHs. 
These simple models illustrate how \halpha~is tracing the star-formation closely, while the UV light is lagging behind $20-40\,{\rm Myrs}$. This causes abrupt changes in the ratio of the SFRs and the \halpha~EW depending on the burstiness of the SFH.
At a constant rate of star-formation (equilibrium condition) the ratio of \halpha~to UV luminosity approaches a value of $-2.1$, the UV SFR becomes comparable to the \halpha~SFR, and the \halpha~EW approaches a time-independent value of roughly $200\,{\rm \AA}$.
In contrast, during the first $\sim50\,{\rm Myrs}$ of a starburst, the \halpha~luminosity and SFR would be enhanced compared to the values measured from UV continuum. The excess depends on the amplitude of the starburst.
Keep in mind that \halpha~EWs lower than the ``equillibrium value'' can only be reached right after a burst or during a phase of strong decreasing star formation activity (Panel C).

We now investigate how model galaxies move on the stellar mass vs. SFR plane depending on their SFH.
For this, we focus on three fiducial galaxies that grow on the main-sequence as parameterized by \citet{SPEAGLE14}
\footnote{We extrapolate this relation to $z=7$, above which the galaxies form stars at a constant rate. This part of the SFH is not critical for our conclusions.} and add some perturbations.
Specifically, model galaxy A experiences a sudden drop in star-formation activity at $z=4.5$. Model galaxy B experiences a burst ($20$ times the mean SFR) at $z=5$ for $100\,{\rm Myrs}$ before its star formation drops. Finally, model galaxy C experiences a burst ($20$ times mean SFR) at $z=4.5$ for $100\,{\rm Myrs}$ but continues to grow along the main-sequence after this event. All galaxies start forming stars at $z=11$ and reach a stellar mass of $\logm = 10.8$, $11.6$, and $11.4$, respectively, at $z\sim4.1$
\footnote{Note that these masses are somewhat arbitrary as they only serve as a normalization parameter.}.

Figure~\ref{fig:mainsequencemodel} shows the redshift evolution of the three model galaxies on the stellar mass vs. SFR plane. In color, we indicate the ratio of \halpha~to UV luminosity (left panel) and the \halpha~EW (right panel).
While growing on the main-sequence, the galaxies increase their SFR more or less exponentially, which results in a time-independent excess in \halpha~SFR of $\sim0.2\,{\rm dex}$ and a steady \halpha~EW of $400-600\,{\rm \AA}$.
During a burst (galaxies B and C), we expect an excess of $\sim0.5\,{\rm dex}$ in \halpha~compared to UV continuum luminosity and an increased \halpha~EW ($>800\,{\rm \AA}$).
On the other hand, if the star formation drops (galaxies A and B), we expect a deficit of $\sim0.8\,{\rm dex}$ in \halpha~compared to UV continuum luminosity (c.f. panel B in Figure~\ref{fig:bursty}). At the same time, the \halpha~EW drops below $200\,{\rm \AA}$.
Galaxy C experiences a starburst at $z\sim4.5$ before it approaches again an equilibrium state with comparable \halpha~and UV luminosities and $\ewha\sim200\,{\rm \AA}$.

In the following, we use the insights from these simple models to understand the recent SFHs of our observed galaxies.

\subsubsection{Interpretation of the observations}
\label{sec:interpretationofobservations}

Figure~\ref{fig:ew45lummass} shows indications of an anti-correlation between \halpha~EW and stellar mass.
Generally, less massive galaxies ($\logm < 10.3$) tend to have higher \halpha~EWs (preferentially $\gtrsim 400\,{\rm \AA}$) and most of them ($>50\%$) show a large excess in \halpha~luminosity and SFR ($0.2-0.5\,{\rm  dex}$ Figure~\ref{fig:mainsequence}). In Section~\ref{sec:isexcessphysical}, we have shown that this excess cannot be explained by uncertainties in our measurements and fiducial parameters only (Figure~\ref{fig:sfrsim}). 
According to our models, a galaxy evolving on the star-forming main-sequence without bursts reaches an \halpha~EW of $\sim600\,{\rm \AA}$. Higher \halpha~EWs and a substantial excess in \halpha~luminosity (or SFR) compared to UV continuum therefore indicate a significantly increased star-formation activity during the past $50-100\,{\rm Myrs}$.
Indeed the observations (Figures~\ref{fig:mainsequence} and~\ref{fig:mainsequenceew}) are broadly consistent with our model galaxy B or C (Figure~\ref{fig:mainsequencemodel}). The large excess in \halpha~luminosity suggests a burst amplitude of $\sim 5-20$ or more (c.f. panels A and B in Figure~\ref{fig:bursty}).
We observe such large \halpha~EWs and excesses in \halpha~luminosity in about $30-50\%$ of our galaxy sample (redshift range between $4 < z < 5$), depending on assumptions. Assuming a $50-100\,{\rm Myrs}$ window during which these conditions can be observed after a burst, we would statistically expect that a galaxy between redshifts $4 < z < 5$ had between $1$ and $4$ bursts with at least $5$ times increased star-formation.

A smaller fraction of galaxies resides below the main-sequence, with low \halpha~EWs ($<200\,{\rm \AA}$) and a deficit in \halpha~compared to UV continuum luminosity expected for a constant star-formation activity. This deficit is even more significant if compared to the expected \halpha~to UV continuum luminosity ratio of a galaxy evolving on the main-sequence. Compared to our model galaxies A and B, suggests that these galaxies must have had a substantial \textit{decrease} in star-formation activity in the past $\sim100\,{\rm Myrs}$. A possible scenario could be that these galaxies had a burst in the past and consumed most of their gas, which ceased their star-formation activity momentarily (model galaxy B). 
At stellar masses of $\logm\sim10-10.5$, it is unlikely that these galaxies will quench entirely, given the low fraction of quiescent galaxies of $<10\%$ at these redshifts \citep[e.g.,][]{PENG10,DAVIDZON17}. It is more likely that their gas reservoirs will be replenished by strong inflows, as it is common in the early universe, and regain their star-formation activity. Such ``compaction, depletion, and replenishment'' cycles are also predicted by current cosmological simulations \citep[e.g.,][]{TACCHELLA16}.

The interpretation of our data at high stellar masses ($\logm > 11$) is more difficult because of the large uncertainties in the dust properties. Assuming that the $f$-factor is redshift dependent and approaches unity at $z=4.5$, massive galaxies are preferentially characterized by low \halpha~EWs ($<200\,{\rm \AA}$) and a deficit in \halpha~luminosity with respect to an equilibrium state in contrast to lower stellar masses. This could indicate that they are on a decreasing branch of star-formation activity. Note that galaxies with stellar masses of $>10^{11}\,{\rm M_{\odot}}$ are expected to have dark matter halo masses of more than $10^{12}\,{\rm M_{\odot}}$ \citep[e.g.,][]{SHETH01,BEHROOZI13}. This prevents infalling gas from cooling and therefore makes these galaxies statistically more prone to cease their star formation \citep{BIRNBOIM03,DEKEL06,KERES09,BIRNBOIM11,WOO13,WOO15,TACCHELLA16,WOO17}. The most massive galaxies with observed decreasing SFH could therefore evolve into the first quiescent galaxies in the early universe at $z>4$.

%%%%%% FIGURE: EMAMI %%%%%%%
\begin{figure}
\includegraphics[width=1.0\columnwidth, angle=0]{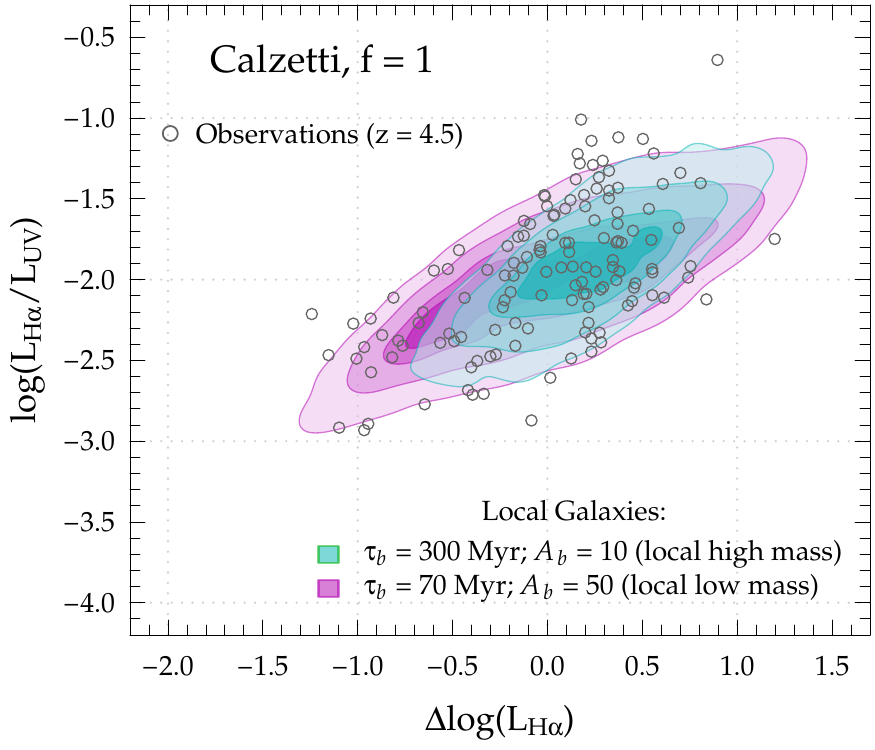}
\caption{Observed $z\sim4.5$ galaxies (open circles) on the $\log({\LHA/\LUV})$ versus $\DLHA$ plane. The distribution of $z\sim4.5$ galaxies is compared to model galaxies with multiple bursts from \citet{EMAMI18} including measurement errors reflecting the ones of our observations. We show a typical local $\logm>9.5$ (turquoise contours) and $\logm < 8.5$ (magenta contours) galaxy, respectively, at local redshifts \citep[][]{EMAMI18}. Note that a local low-mass galaxy model fits well to our observations at high redshifts, suggesting similar burst properties in the population of low-mass local and massive high-redshift galaxies.
\label{fig:emami}}
\end{figure}
%%%%%%%%%%%%%%%%%%%%%%%%%%%%%

\subsubsection{Comparison to galaxies at lower redshifts}

How do our findings compare to galaxies at later cosmic times?
While galaxies of stellar masses $\logm \sim 10.5-11.0$ at late cosmic times likely change their star-formation activity only smoothly over long timescales, the SFH of low-mass ($\logm < 9$) galaxies is expected to be variable on short timescales \citep[usually less than $\sim100\,{\rm Myrs}$, ][]{WUYTS11,WEISZ12,SHEN14,DOMINGUEZ15,CONROY16,GUO16,SHIVAEI16,EMAMI18,BROUSSARD19}. Possible reasons are feedback mechanisms such as from AGN or Supernovae \citep[e.g.,][]{BOSELLI09,HOPKINS14,SPARRE17}, that affect the star-formation activity of low-mass galaxies with shallow gravitational potentials.
On the other hand, changes in the star-formation activity of more massive galaxies are likely dominated by gas inflows and galaxy-galaxy interactions that change star-formation on timescales of several $100\,{\rm Myrs}$ \citep[][]{SPARRE15,TACCHELLA16}.

In our sample at $z\sim4.5$ we observe large ratios of \halpha~to UV luminosity (or SFR) as well as high \halpha~EWs. A large fraction of galaxies at $\logm>10$ show such properties, which suggest a substantial variability in star-formation activity. According to our models, these changes happen on short timescales with large amplitudes of factors of $5-10$ or more. This indicates that the star-formation activity at high redshifts across a large mass range is similar to low-mass galaxies in the local universe. The reasons of the burstiness have yet to be revealed. Galaxy-galaxy interactions, gas fractions, and gas inflow rates are expected to be higher at early cosmic times \citep{SILVERMAN15,SCOVILLE16}, hence could cause large variations in star-formation activity \citep[it is also expected in a theoretical picture, see][]{FAUCHER18}.

In the following, we discuss this more qualitatively.
Figure~\ref{fig:emami} relates $\log({\LHA/\LUV})$ with $\DLHA$ for the case of Calzetti dust attenuation and $f=1$\footnote{We only show the case of Calzetti dust attenuation and $f=1$ (likely valid for high redshift galaxies, Section~\ref{sec:ffactor}), however, other previously discussed values do not significantly change the results of this very qualitative analysis.}.
Thereby, $\DLHA$ is defined as the logarithmic deviation of a galaxy from the average $\LHA$ vs. stellar mass relation \citep{EMAMI18}.
The contours show the best-fit models to local galaxies with stellar masses of $\logm<8.5$ (magenta) and $\logm>9.5$ (turquoise) from \citet{EMAMI18}. The models are defined by multiple bursts of star formation with e-folding time of ($\tau_b$) and amplitude of ($A_b$).
As discussed in their paper \citep[see also][]{WEISZ12}, low-mass galaxies are characterized by shorter bursts with higher amplitude, while the opposite is true for massive galaxies, consistent with the current theoretical picture.
Our observations at $z\sim4.5$ (empty gray circles) coincide well with the parameter space covered by local low-mass galaxies (with statistically a short burst duration ($\tau_b<70\,{\rm Myr}$) and high amplitude ($A_b=50$)). Specifically, their distribution shows a larger \textit{extent} compared to local galaxies of similar stellar mass. This indicates that high-redshift galaxies show statistically a more bursty SFH than same-mass galaxies at low redshifts.
Note that besides the changes in the SFR, uncertainties in \halpha~and UV continuum measurements affect the $\log({\LHA/\LUV})-\DLHA$ distribution. We ``widened'' the models of the low-redshift galaxies to take the effect of the larger measurement uncertainties at $z\sim4.5$ into account, in order to obtain a fair comparison between models and observations. Furthermore, we make the reader aware that the sample used in \citet{EMAMI18} is naturally biased against interacting galaxies, which could underestimate the burstiness of massive galaxies.

Since Emami et al. used a relatively simple SFH in their models, we compare our results to a more stochastic representation of burstiness. Specifically, we use the burst models introduced in \citet{CAPLAR19}, where burstiness is defined stochastically by a power-law slope $\alpha_{\rm pl}$ (fixed here to $\alpha_{\rm pl}=2$) and decorrelation timescale $\tau_{\rm break}$. We apply this characterization of bursts to the galaxy evolution model in \citet{TACCHELLA18} to generate more realistic SFHs. Assuming a $\tau_{\rm break} < 100\,{\rm Myrs}$ leads in a comparable relation on the $\log({\LHA/\LUV})$ versus $\DLHA$ plane as our observed galaxies. This result is qualitatively in agreement with our conclusions above of a substantially bursty SFH at high-$z$.

%%%%%% FIGURE: IONIZING PHOTON PRODUCTION EFFICIENCY %%%%%%%
\begin{figure}
\includegraphics[width=1.0\columnwidth, angle=0]{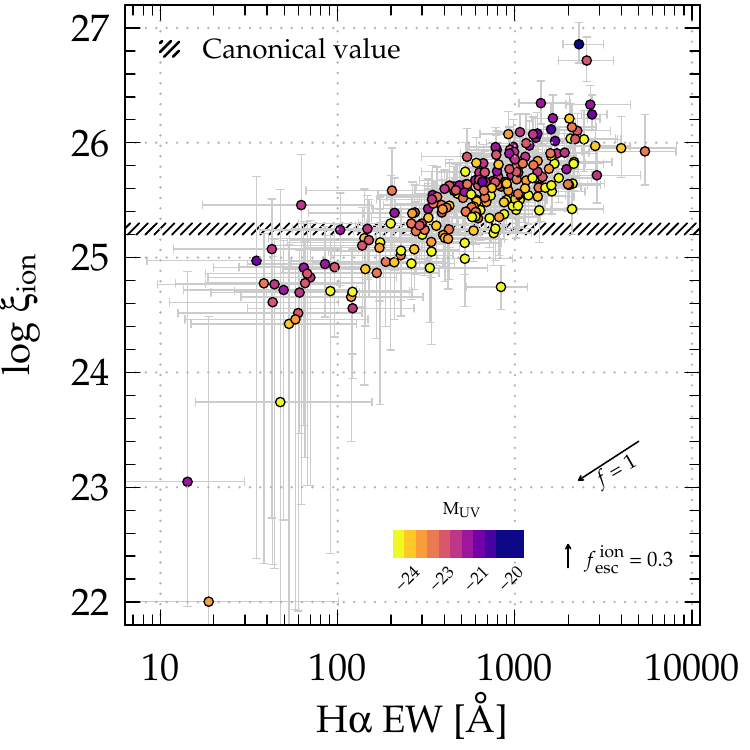}
\caption{The ionizing photon production efficiency (assuming \fion$=0.1$) as a function of \halpha~EW and absolute UV magnitude (color-coded), Uncertainties due to the $f$-factor and the escape fraction of ionizing continuum (\fion) are indicated with arrows. Our sample spans a large range in \xiion with a median of $\log(\xi_{\rm ion}) \sim 25.5$. This is $\sim0.3$ higher then the typically assumed canonical value \citep{BOUWENS16}. The steep correlation between \xiion~and \halpha~EW (i.e. sSFR) has to be taken into account when using \xiion~to model the contribution of galaxies to reionization.
\label{fig:xi}}
\end{figure}
%%%%%%%%%%%%%%%%%%%%%%%%%%%%%

%%%%%% FIGURE: LYA ESCAPE FRACTION %%%%%%%
\begin{figure}
\includegraphics[width=0.95\columnwidth, angle=0]{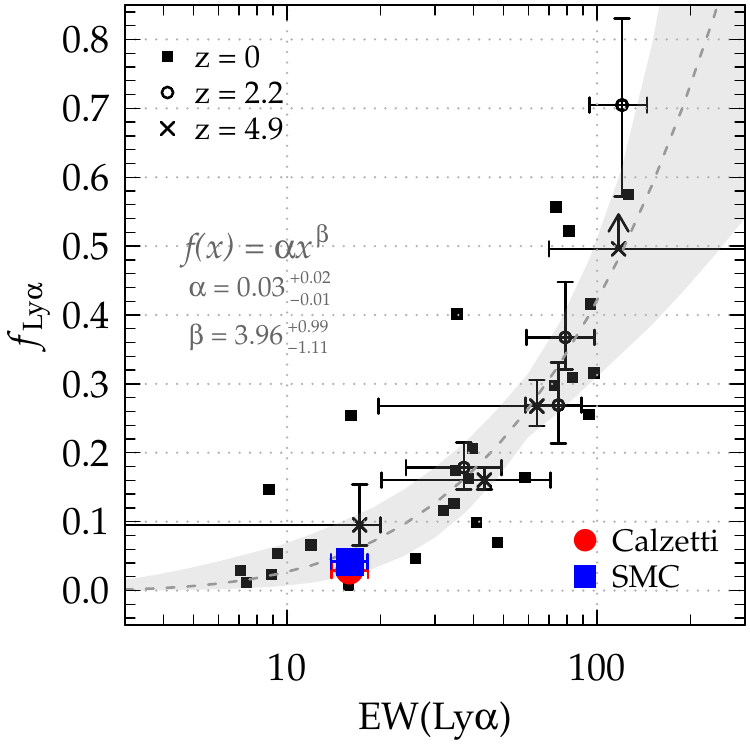}
\caption{Our estimate of the \lya~escape fraction at $z\sim4.5$ (commonly $<15\%$) compared to values in the literature. These include a compilation of local galaxies \citep[squares,][]{CARDAMONE09,HECKMAN05,OVERZIER09,HAYES14,OSTLIN14,HENRY15,YANG16} as well as Lyman-$\alpha$ emitters at $z\sim2.2$ \citep[open points,][]{SOBRAL17} and $z\sim4.9$ \citep[crosses,][]{HARIKANE18}. Our results are consistent with a global and redshift-independent relation between \flya~and the \lya~EW (gray band, $95-$percentiles of fit, see Equation~\ref{eq:lyaescape}). The LAE samples are narrw-band selected, hence probing a galaxy population with larger \lya~EWs. \label{fig:lyaescape}}
\end{figure}
%%%%%%%%%%%%%%%%%%%%%%%%%%%%%

\subsection{Expected Range in the Ionizing Photon Production Efficiency due to Bursty Star-Formation Activity}

The efficiency of ionizing photon production (\xiion, in ${\rm Hz\,erg^{-1}}$) depends on the output of ionizing photons, hence directly relates to the \halpha~luminosity. In a broader picture, \xiion~is, together with the escape fraction of ionizing continuum radiation, an important quantity to understand the reionization of the early universe by galaxies \citep{ROBERTSON15,FAISST16c, BOUWENS16b,HARIKANE18,SHIVAEI18,LAM19}. Observationally it is found that $\log(\xi_{\rm ion}) \sim25.2$, which is typically assumed as the canonical value \citep[e.g.,][]{BOUWENS16}. However, as the \halpha~to UV luminosity varies depending on the recent SFH, a significant scatter around this value is expected. Our large sample of galaxies at $z\sim4.5$ with robust \halpha~measurements enables us to quantify this scatter.

To calculate \xiion, we follow the recipe used in \citet{HARIKANE18},
\begin{equation}
    \xi_{\rm ion} = \frac{N({\rm H^0})}{L_{\rm UV}},
\end{equation}
where $L_{\rm UV}$ is the UV luminosity and $N({\rm H^0})$ is the production rate of ionizing photons (in ${\rm s}^{-1}$). The latter can be calculated using the conversation factor derived in \citet{LEITHERERHECKMAN95} as
\begin{equation}
    L^{\rm int}_{\rm H\alpha} = 1.36\times10^{-12}\,(1 - f^{\rm ion}_{\rm esc})\, N({\rm H^0}),
\end{equation}
where \fion~is the \textit{internal} ionizing photon escape fraction.
In the following, we assume a fiducial value of \fion$\sim0.1$, which is reasonable given the values found in the current literature \citep[see][]{HARIKANE18}.
Figure~\ref{fig:xi} shows \xiion~as a function of \halpha~EW and absolute UV magnitude (color-coded). The uncertainties introduced by the $f$-factor and \fion~(shown as arrows) do not impact significantly the following results (note that \xiion~and \halpha~EW are equally affected by $f$). The correlation between these quantities is trivial as $\xi_{\rm ion} \propto L_{\rm H\alpha}/L_{\rm UV}$, which is roughly proportional to the \halpha~EW assuming the output of rest-frame UV and optical light is well correlated (which is mostly true for galaxies with recent star-formation). More interesting is the range in \xiion, from $\log(\xi_{\rm ion}) \sim24.5-26.4$ with a median of $\sim 25.5$, which is $\sim0.3$ dex above the typically used canonical value. Some of the most massive galaxies (with suppressed \halpha~EWs, Figure~\ref{fig:ew45lummass}) show values significantly below this range, the opposite is true for highly star-forming galaxies at low stellar masses. The correlation between \xiion~and \halpha~EW (i.e. sSFR) has to be taken into account when using \xiion~to model the contribution of galaxies to reionization.
As reported in other studies, we recover a trend of higher \xiion~with fainter absolute UV magnitude \citep{BOUWENS16}. 

A quantity that is related to the production and escape of ionizing photons is the average \lya~escape fraction (\flya), which can be directly constrained with our sample by
\begin{equation}
    f_{\rm Ly\alpha} = \frac{L^{\rm obs}_{\rm Ly\alpha}}{L^{\rm int}_{\rm Ly\alpha}} = \frac{L^{\rm obs}_{\rm Ly\alpha}}{8.7\,L^{\rm int}_{\rm H\alpha}}
\end{equation}
assuming case B recombination \citep{BROCKLEHURST71} and using the available measurements of \lya~from rest-UV Keck/DEIMOS spectroscopy \citep{MALLERY12,HASINGER18}.
The estimated \flya~for our sample is less than $10\%$ (Figure~\ref{fig:lyaescape}), which is significantly lower than reported for Ly$\alpha$ emitters at the same redshift \citep{HARIKANE18}. However, taking into account the $\sim8$ times larger \lya~EWs measured for the LAE sample, our measurements fit well in a global, redshift-independent relation between \flya~and \lya~EW that can be described as
\begin{equation}\label{eq:lyaescape}
    \log f_{\rm Ly\alpha} = \alpha\,\left[ \log {\rm EW(Ly\alpha)} \right]^\beta.
\end{equation}
A fit to all the literature data (Figure~\ref{fig:lyaescape}) results in $\alpha = 0.03^{+0.02}_{-0.01}$ and $\beta = 3.96^{+0.99}_{-1.11}$.

%%%%%%%%%%%%%%%%%%%%%%%%%%%%%
%%%		CONCLUSIONS		  %%%
%%%%%%%%%%%%%%%%%%%%%%%%%%%%%
\section{Summary and Conclusions} \label{sec:end}

The goal of this work is to investigate the burstiness of recent star-formation at high redshifts using \halpha~and UV continuum emission measurements.

For this, we measure the \halpha~luminosities and EWs for a mass-complete sample of $221$ galaxies with $\logm > 9.7$ at $3.9 < z < 4.9$. In this redshift range, \halpha~emission can be measured using the Spitzer \iracA~color without contamination by other strong optical emission lines. The sample includes galaxies with spectroscopic and robust photometric redshifts. Our method to measure \halpha~emission from Spitzer colors does properly propagate uncertainties in stellar population age and dust attenuation, which dominate the slope of the optical continuum (Section~\ref{sec:emlinfit}).

It is worth noting that the largely unconstrained differential dust attenuation between stars and nebular regions ($f$-factor) in high-redshift galaxies causes the largest uncertainties in our analysis. This is particularly true for the massive dust-rich galaxies in our sample. Recent studies at $z\sim2$ suggest that $f$ may be closer to unity hence may significantly deviate from the standard $0.44$ measured in local galaxies due to the different ISM conditions. In Section~\ref{sec:ffactor}, we test this scenario with local galaxies that have increased \halpha~EW analogous to high-$z$ galaxies. We find indeed an  increase in $f$ with \halpha~EW or sSFR (closely related to star-formation density). This motivates the use of a high $f$-factor for high-redshift galaxies.

Summarizing the main results (Section~\ref{sec:results}), we find a significant scatter between \halpha~and UV continuum derived luminosities and SFRs (Figure~\ref{fig:lumsfrcomparison}). About half of the galaxies show an excess (more than a factor of $3$) in \halpha~SFR and luminosity compared to what would be expected for a constant (i.e., smooth) SFH. Observational uncertainties and model assumptions alone cannot explain this (Figure~\ref{fig:sfrsim}).
The fraction of such galaxies is less at higher stellar masses. In addition, we find indications of an anti-correlation between \halpha~EW and stellar mass (Figure~\ref{fig:ew45lummass}).

A comparison to simple models shows that the excess in \halpha~luminosity and SFR may be due to a recent increase in star-formation activity (Figure~\ref{fig:bursty}). Specifically, an increase $5-20$ times over the mean SFR must have happened over the past $\sim 50\,{\rm Myrs}$. We note that a continuously increasing SFH (which is expected for galaxies growing on the main-sequence) would not result in such high \halpha~luminosities or EWs. Hence, an additional starburst is necessary on top of the underlying smooth star formation. Such a model also explains the large scatter (up to $1\,{\rm dex}$) of the \halpha-derived SFR vs. mass main-sequence compared to the UV derived main-sequence (Figures~\ref{fig:mainsequence}, \ref{fig:mainsequenceew}, and~\ref{fig:mainsequencemodel}).
From statistical arguments, we estimate that a galaxy has experienced on average $1-4$ major bursts ($>5$ times increased star-formation) between $4 < z < 5$.

We find tentative differences as a function of stellar mass. Specifically, massive galaxies ($\logm > 11$) have a lower average \halpha~EW (Figure~\ref{fig:ew45lummass}). Furthermore, the fraction of massive galaxies with excess \halpha~emission (compared to values expected for a constant SFR) is lower compared to lower masses and consistent with no excess or even a deficit when including measurement uncertainties (Figure~\ref{fig:sfrsim}).
This hints toward a decreasing star-formation activity in these galaxies (Figure~\ref{fig:mainsequencemodel}). With current measures, it is not possible to tell whether these galaxies are becoming quiescent or are only experiencing a temporary decrease in star-formation. However, the most massive $\logm>11$ galaxies in our sample are likely embedded in dark matter halos of masses of $>10^{12}\Msol$ and therefore have an increased probability to ceases their star formation in the near future.

In the local universe, galaxies at low stellar masses $\logm < 8.5$ are found to have a bursty SFH, while more massive galaxies evolve more smoothly. Our analysis suggests that this is not the case a high redshifts; galaxies at $z\sim4.5$ show signs of bursts across a large stellar mass range, even at $\logm>10$, similar to local low-mass galaxies (Figure~\ref{fig:emami}). This goes along with the high star-formation activity and fast mass build-up of these galaxies, driven by high gas fractions, gas inflows, and possibly galaxy-galaxy interactions.

Finally, the bursty nature of star-formation has an impact on the ionizing photon production efficiency. We report a median $\log(\xi_{\rm ion})$ of $25.5$, which is $0.3\,{\rm dex}$ above the canonical value typically adopted. The range in $\log(\xi_{\rm ion})$ in our sample due to variations in \halpha~emission is large, spanning $24.5$ to $26.4$ (Figure~\ref{fig:xi}).

The power of using Spitzer broad-band photometry to derive constraints on the SFH of galaxies and hence to study their growth in the early universe is evident. The large area of current and ongoing Spitzer surveys probe a different parameter space than JWST, which will provide accurate \halpha~emission measurements from high-resolution spectroscopy for subsets of galaxies $-$ a fraction of what is observed with Spitzer.
JWST will be especially important to constrain the differential dust attenuation (the main uncertainty in this analysis) at these redshifts by measuring the ratio of Balmer lines.
Furthermore, changes in metallicity and IMF may affect the \halpha~to UV flux ratio. Although we do not expect these to have a significant impact, future constraints on these quantities from spectroscopic observations with JWST will certainly reduce uncertainties.

\acknowledgements
\textit{Acknowledgements:} We thank the referee for their valuable comments which improved this paper as well as J. Kartaltepe, S. Toft, and N. Scoville and  for useful and fruitful discussions.
N.E. acknowledges support from Program number 13905 provided by NASA through a grant from the Space Telescope Science Institute, which is operated by the Association of Universities for Research in Astronomy, Incorporated, under NASA contract NAS5-26555.
S.T. is supported by the Smithsonian Astrophysical Observatory through the CfA Fellowship.
This work is based on observations and archival data made with the \textit{Spitzer Space Telescope}, which is operated by the Jet Propulsion Laboratory, California Institute of Technology, under a contract with NASA, along with archival data from the NASA/ESA Hubble Space Telescope.
This research made also use of the NASA/IPAC Infrared Science Archive (IRSA), which is operated by the Jet Propulsion Laboratory, California Institute of Technology, under contract with the National Aeronautics and Space Administration.
In parts based on data products from observations made with ESO Telescopes at the La Silla Paranal Observatory under ESO programme ID 179.A-2005 and on data products produced by TERAPIX and the Cambridge Astronomy Survey Unit on behalf of the UltraVISTA consortium.
Based on data obtained with the European Southern Observatory Very Large Telescope, Paranal, Chile, under Large Program 185.A-0791, and made available by the VUDS team at the CESAM data center, Laboratoire d'Astrophysique de Marseille, France.
Furthermore, this work is based on data from the W.M. Keck Observatory and the Canada-France-Hawaii Telescope, as well as collected at the Subaru Telescope and retrieved from the HSC data archive system, which is operated by the Subaru Telescope and Astronomy Data Center at the National Astronomical Observatory of Japan.
The authors wish to recognize and acknowledge the very significant cultural role and reverence that the summit of Mauna Kea has always had within the indigenous Hawaiian community. We are most fortunate to have the opportunity to conduct observations from this mountain.
Finally, we would also like to recognize the contributions from all of the members of the COSMOS Team who helped in obtaining and reducing the large amount of multi-wavelength data that are now publicly available through IRSA at \url{http://irsa.ipac.caltech.edu/Missions/cosmos.html}.

\clearpage
\newpage
\appendix

\section{A machine learning approach to assess photometric contamination}
\label{sec:appendixA}

%%%%%% FIGURE: T-SNE %%%%%%%
\begin{figure*}[b!]
\includegraphics[width=1.0\columnwidth, angle=0]{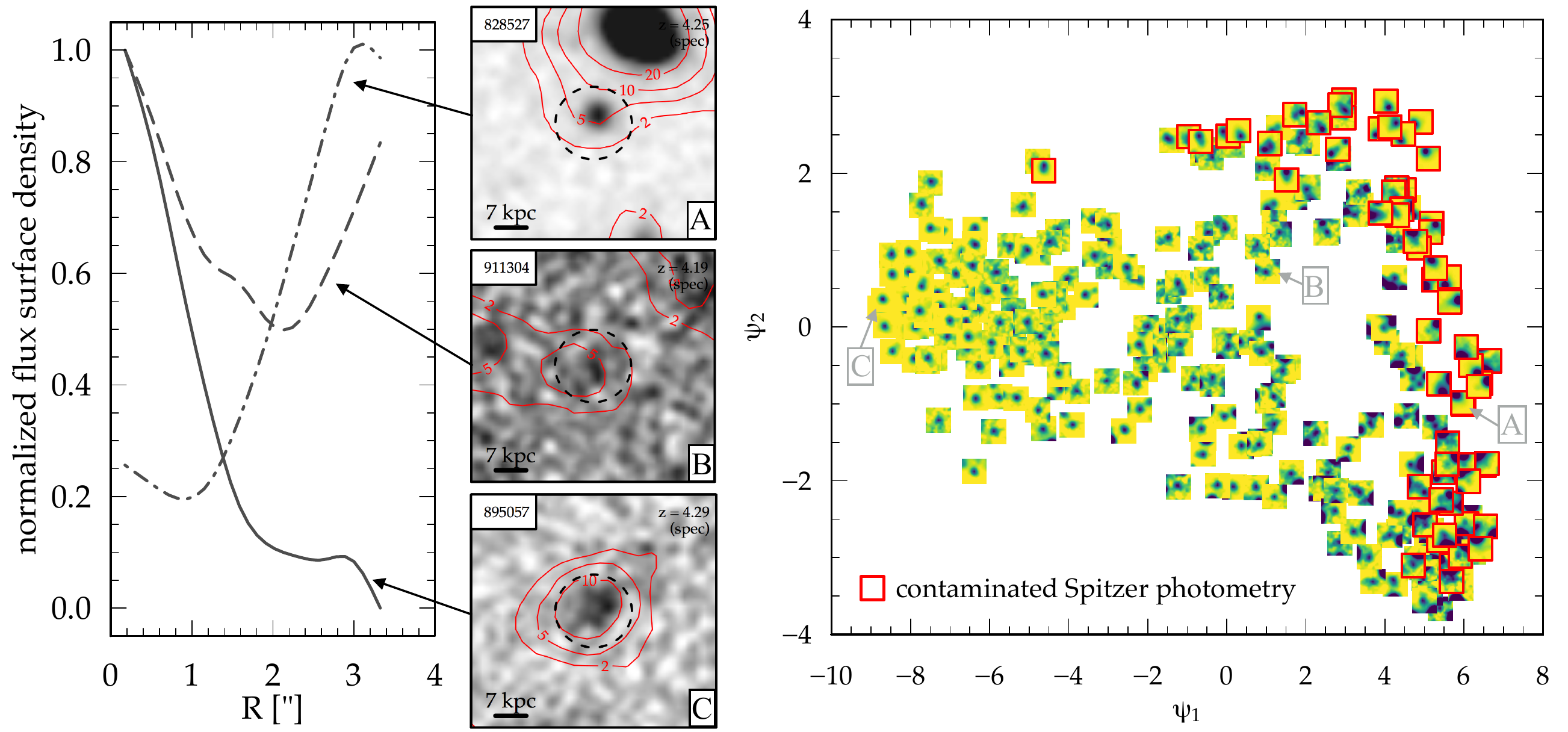}
\caption{Quantitative assessment of photometric contamination using unsupervised machine learning. \textit{Left:} Three examples of surface-brightness curves measured in ten non-overlapping shells out to $4\arcsec$ centered on our galaxies. The galaxies A through C are representative of galaxies whose Spitzer fluxes are strongly, weakly, and not at all contaminated by neighboring galaxies, respectively. The curves are normalize to one at their maximum. The cutouts show UltraVISTA $K$-band images with Spitzer $3.6\,{\rm \mu m}$ contours overlaid. The circle has a radius of $2.5\arcsec$ and the cutouts are $7\arcsec$ across.
\textit{Right:} Grouping of our galaxies (shown are Spitzer $3.6\,{\rm \mu m}$ cutouts) derived by the \texttt{t-SNE} algorithm applied to the surface-brightness curves. Galaxies classified visually to have contaminated photometry (i.e., the training sample) are shown with a red box. The \texttt{t-SNE} algorithm separates them naturally to the right side in the ($\psi_1,\psi_2$) space. \label{fig:tsne}}
\end{figure*}
%%%%%%%%%%%%%%%%%%%%%%%%%%%%%

In Section~\ref{sec:spitzercontamination}, we have rejected galaxies whose Spitzer photometry is contaminated by bright nearby sources using imaging data at higher spatial resolution combined with visual inspection. For large amounts of data, such an approach clearly becomes unfeasible. In these cases, a machine learning approach can result in statistically similar results as by human inspection but at a fraction of the time.

Here, we demonstrate how machine learning can be used to quantify the contamination of the photometry of galaxies. We use our visual classification to test this method.
Specifically, we use the \textit{t-Distributed Stochastic Neighbor Embedding} algorithm \citep[\texttt{t-SNE}\footnote{Here we use the \textit{R} package \texttt{Rtsne}. More information at \url{https://lvdmaaten.github.io/tsne/}.},][]{MAATEN08}, which is an unsupervised machine learning algorithm used for dimensionality reduction of large datasets.

In order to apply \texttt{t-SNE} to our problem, we first have to derive estimators for the contamination of photometry and normalize them correctly.
For this, we measure the average Spitzer flux in ten non-overlapping shells with radius of $0.4\arcsec$, out to $4\arcsec$ centered on the galaxies.
The shapes of these surface-brightness curves should directly characterize the amount of contamination from neighboring galaxies. In contrast to simply counting the number of companions (as done in Section~\ref{sec:spitzercontamination}), this approach takes into account the shape and brightness of the target and contaminating galaxies. Examples of such profiles and UltraVISTA $K$-band image cutouts are shown in the left panel of Figure~\ref{fig:tsne}).

Using machine learning, we explore the parameter space of different surface-brightness profiles. \texttt{t-SNE} automatically will group similar profiles without a training sample in hand. The training sample (i.e., our visual classification) is used to identify the different groups and to assign them a level of contamination.

The right panel of Figure~\ref{fig:tsne} shows the output of \texttt{t-SNE} on a parameter space spanned by two dimensionless vectors $\psi_1$ and $\psi_2$ that are introduced by the algorithm.
The cutouts show the Spitzer $3.6\,{\rm \mu m}$ images of the $284$ galaxies and the red boxes mark the galaxies that we have found to be contaminated by neighboring galaxies based on our previous visual inspection of the $K$-band images. The \texttt{t-SNE} algorithm naturally gathers these galaxies on the right side in the ($\psi_1,\psi_2$) parameter space. The distance in 2-dimensional space reflects directly the similarities in the light profiles. In the case discussed here, there is a gradient from left to right from low to high contamination. The parameter space can be interpolated to assign a continuous probability of contamination to each of our galaxies, for example by using a k-nearest-neighbor algorithm.
In Figure~\ref{fig:sample}, we indicate the galaxies at $\logm > 10.5$ with a contamination probability $>0$ in red (note that the galaxies that were visually classified to be contaminated are not shown in that figure). The probability is computed as the fraction of visually classified contaminated galaxies among the 5-nearest-neighbors at a given position in ($\psi_1,\psi_2$) parameter space.
Using this definition and assuming that a galaxy's photometry is contaminated if this probability is $\geq0.5$, we can compute the performance of this method by comparing the results to our visual classification (thereby assuming it represents the truth). We obtain a purity (i.e., precision) of $75\%$ and a completeness (i.e., recall) of $70\%$. We note that these numbers would likely improve if tested on a large sample.

\section{How to correct for different $f$-factors}\label{sec:appendixffactor}

\begin{figure}[b!]
\center
\includegraphics[width=0.5\columnwidth, angle=0]{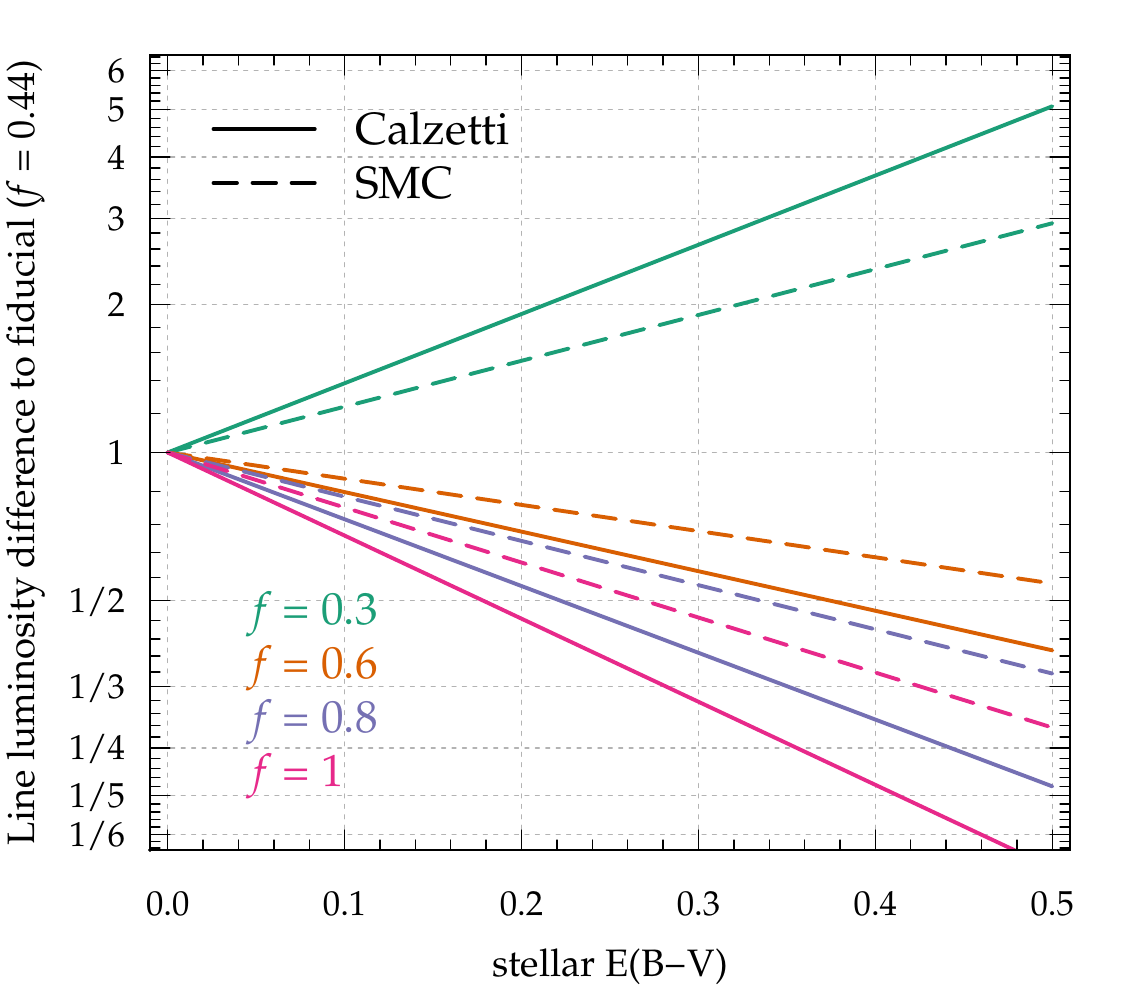}
\caption{Impact of deviations of the differential dust attenuation ($f$-factor) from $0.44$ \citep[as measured in local starburst galaxies][]{CALZETTI00} on quantities linearly related to \halpha~luminosity. For stellar \ebmv~of less than $0.2\,{\rm mag}$, the correction is less than a factor of two for a value of $f=0.8$ (expected at higher redshifts).   \label{fig:febmv}}
\end{figure}

The correction of \halpha~EWs and luminosities, as well as other linearly depending parameters, for deviations from our fiducial $f$-factor is straight forward by applying the multiplicative factor, $\phi$, defined as
\begin{equation}
    \phi(f_1,f_2;\lambda)\vert_{E_{s}{\rm (B-V)}} = 10^{0.4 E_{s}{\rm (B-V)} k(\lambda)  \left[\frac{1}{f_2} - \frac{1}{f_1} \right]},
\end{equation}
where $\ebmvs$ is the \textit{stellar} dust attenuation, $f_{i}$ are the $f$-factors, and $k(\lambda)$ describes the reddening curve.
Figure~\ref{fig:febmv} shows that $f$-factors deviating from $0.44$ can have a significant impact on the measured \halpha~luminosities. For example, assuming $f=0.8$ would decrease the \halpha~luminosities and EWs by roughly a factor two for a moderate dust attenuation of $\ebmvs = 0.2$ assuming a \citet{CALZETTI00} dust attenuation law.
(Note that throughout this paper all dust attenuation values are given in their stellar values unless noted otherwise.)

%%%%%% FIGURE: Febmv %%%%%%%
%%%%%%%%%%%%%%%%%%%%%%%%%%%%%

\section{Detailed Study of Biases in the measurement of \halpha~emission} \label{sec:biases}

In the following, we provide a detailed study of the \textit{(i)} systematic model-based biases of our method in the case of ideal observations (i.e., no photometric noise) and \textit{(ii)} observation-based biases that occur due to the flux limits of the Spitzer surveys.

%%%%%% FIGURE: TEST-METALLICITY %%%%%%%
\begin{figure}[t!]
\includegraphics[width=1.0\columnwidth, angle=0]{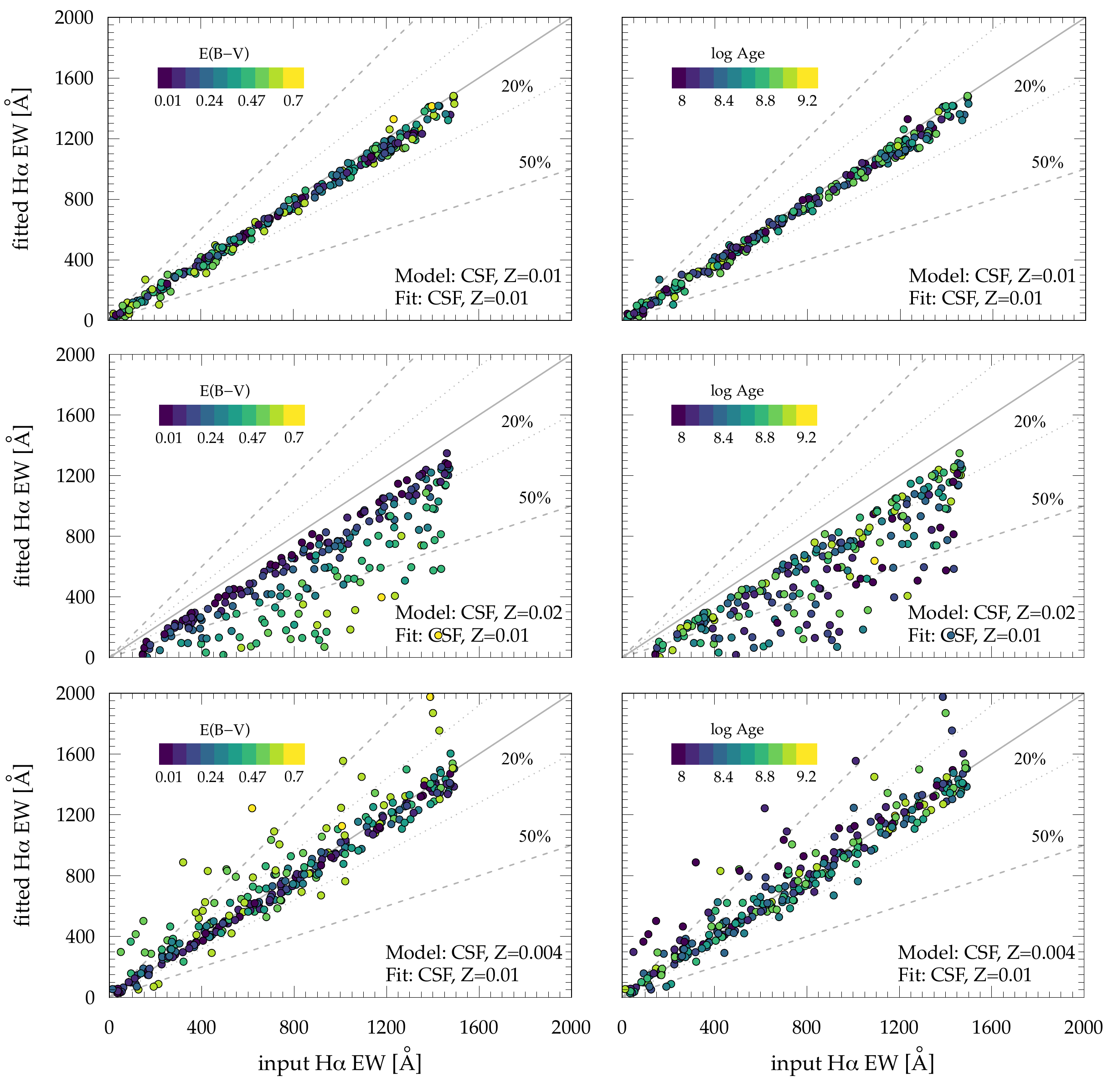}
\caption{Systematic biases in the \halpha~EW measurements if assuming a half-solar metallicity (``fit'') for different underlying (true) metallicities (``model'') as a function of dust (left panels) and age (right panels). A constant SFH is assumed in all cases. The biases mostly depends on $\ebmvs$~of the galaxies but not on their stellar population ages.\label{fig:fittestmetallicity}}
\end{figure}
%%%%%%%%%%%%%%%%%%%%%%%%%%%%%

\subsection{Biases due to model assumptions}

To study the impact of different model assumptions on the derived \halpha~EWs and luminosities, we create $15\,000$ mock galaxies between redshifts $4.0 < z < 5.0$ with stellar continua based on the \citet{BRUZUALCHARLOT03} library. To each galaxy, we assign a SFH (constant or exponentially decreasing with different $\tau$), age (only constrained by the age of the Universe), stellar metallicity ($0.2$, $0.5$, or $1$ times solar), and stellar dust attenuation ($\ebmvs$ $=0-0.7$) fixed to \citet{CALZETTI00}. We assume an $f$-factor of $0.44$. The \halpha~line emission is set by the \halpha~EW (chosen randomly between rest-frame $10\,{\rm \AA}$ and $5000\,{\rm \AA}$) and the stellar continuum.
We add other emission lines in an identical way as described in Section~\ref{sec:emlinfitassumptions}.
Each galaxy SED including emission lines is then convolved with the Spitzer filter transmission curves to obtain the final photometry (we do not include photometric errors at this point).

We use our method to measure the \halpha~emission of the mock galaxies from their Spitzer colors, assuming a fiducial optical continuum based on a constant SFH and half-solar stellar metallicity as well as a Calzetti dust attenuation law. For now, we assume that we know the most likely $\ebmvs$~and age of the galaxies, hence we use a 2-dimensional Gaussian PDF centered on these values with $\sigma_{\log({\rm age})} = 0.2$ and $\sigma_{\rm \ebmvs} = 0.05$.

\begin{figure}
\includegraphics[width=1.0\columnwidth, angle=0]{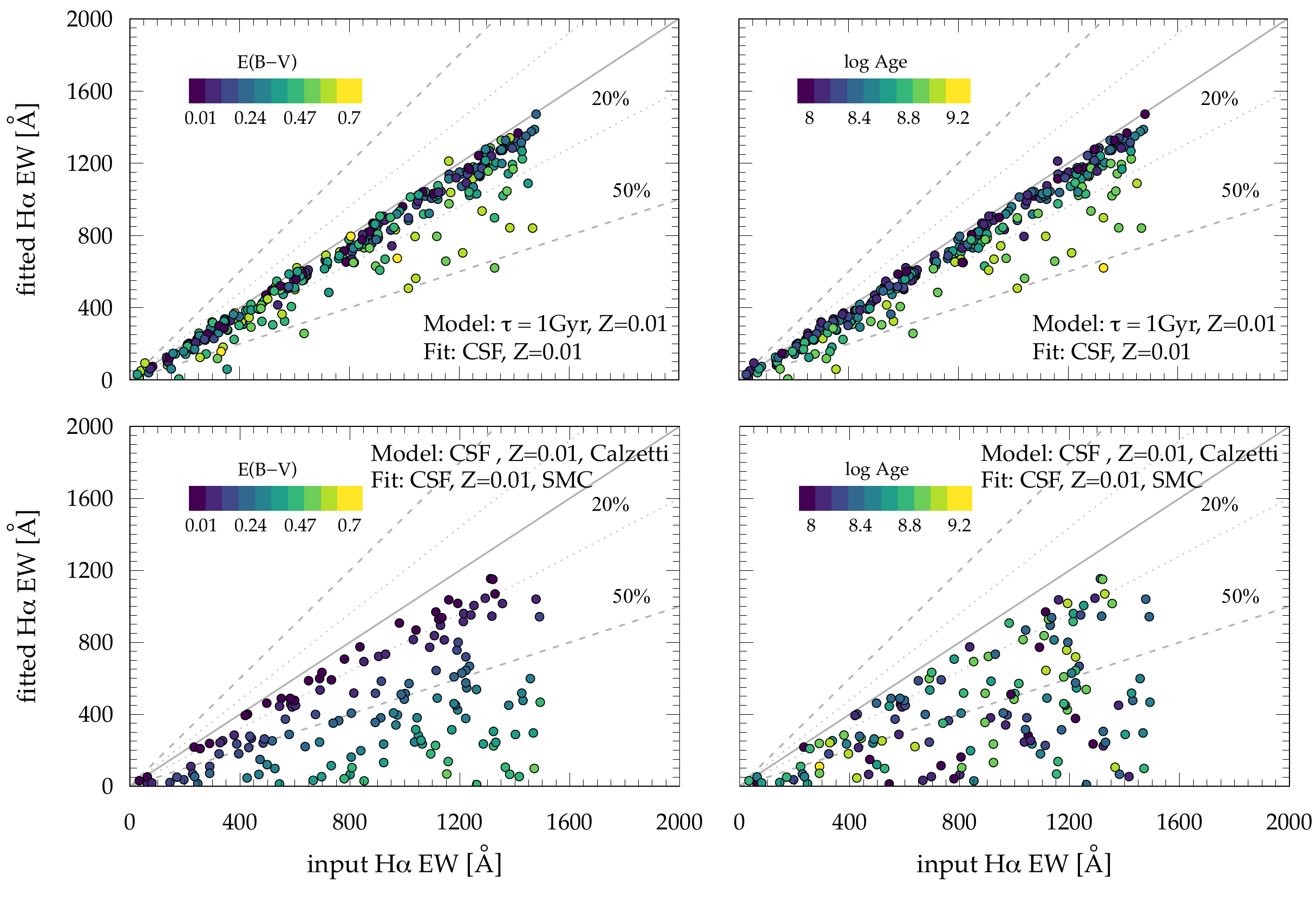}
\caption{Same as Figure~\ref{fig:fittestmetallicity} but for a different SFH or dust attenuation at a fixed metallicity. While the former bias is negligible (within $20\%$), the latter bias is strongly correlated with $\ebmvs$~and can become severe ($>50\%$) for $\ebmvs\gtrsim 0.25$.  \label{fig:fittestsfhdust}}
\end{figure}

Figures~\ref{fig:fittestmetallicity} and \ref{fig:fittestsfhdust} summarize the results of these tests by comparing the input to the output \halpha~EWs. Each of the horizontal pairs of panels show the results for a set of mock galaxies with different SFH and metallicity.
The points are color-coded by $\ebmvs$~(left panels) and stellar population age (right panels). For the lower panels in Figure~\ref{fig:fittestsfhdust}, we assume an SMC dust extinction law for the fitting.

%%%%%% FIGURE: TEST-SFH-DUST %%%%%%%
%%%%%%%%%%%%%%%%%%%%%%%%%%%%%

%%%%%% FIGURE: Dependence on S/N %%%%%%%
\begin{figure*}
\includegraphics[width=1.0\columnwidth, angle=0]{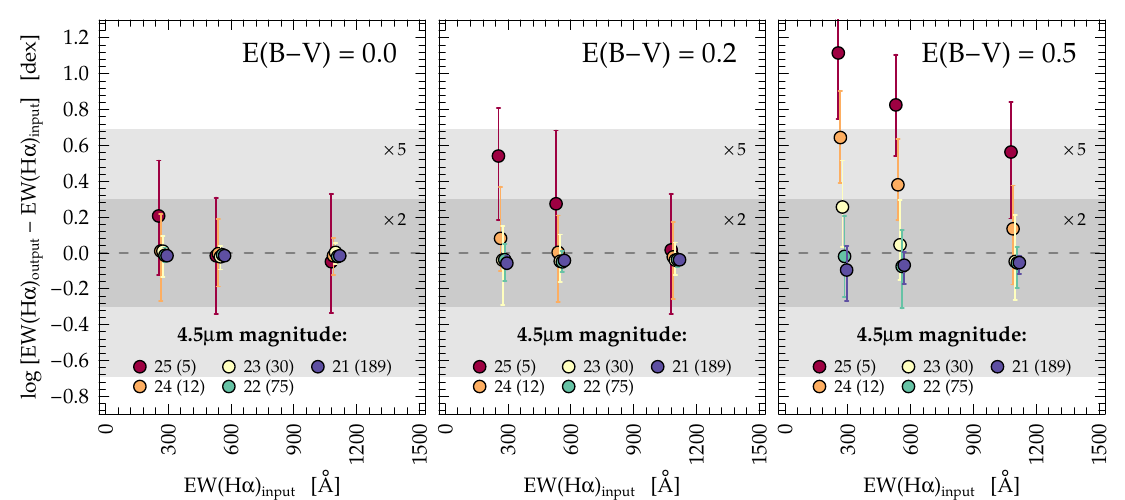}
\caption{Biases in the \halpha~EW measurement due to photometric noise for three bins of EW at $z=4.5$ and increasing $\ebmvs$~(panels from left to right). The symbols show median fitted EWs for three different observed $4.5\,{\rm \mu m}$ magnitudes (S/N in given parenthesis, S/N$=5$ cut is applied to sample). The error bars show the $1\sigma$ scatter and the gray wedges show $20\%$ and $50\%$ deviations from the input (true) values. There are no significant biases in the \halpha~EW measurements for $\ebmvs<0.2$ at S/N $>5$. At higher dust attenuation, we would expect severe biases for faint galaxies, however, as shown in Figure~\ref{fig:sample0}, galaxies with $\ebmvs>0.2$ have observed magnitudes brighter than $24\,{\rm AB}$ at $4.5\,{\rm \mu m}$ (S/N$>12$), hence we to not expect severe biases in their measurements. \label{fig:fittestsSN}}
\end{figure*}
%%%%%%%%%%%%%%%%%%%%%%%%%%%%%

As expected, our method recovers the \halpha~EWs accurately with minimal scatter or biases with the correct assumption of SFH, metallicity, and dust attenuation curve (upper panels, Figure~\ref{fig:fittestmetallicity}).
An underestimation of the metallicity (here factor of 2) results in an underestimation of the \halpha~EWs and vice versa (middle and lower panels of Figure~\ref{fig:fittestmetallicity}).
The biases due to metallicity are generally at a level of less than $20\%$ for galaxies with $\ebmvs\lesssim0.3$. However, for larger dust attenuation, these biases are more severe and can reach $50\%$ or up to a factor of two in the worst cases.
Changing the SFH from constant to exponentially declining (as seen in quiescent galaxies) at fixed metallicity has little impact on the results (top panels of Figure~\ref{fig:fittestsfhdust}).
On the other hand, assuming a steeper dust attenuation curve than the true (in our case SMC instead of Calzetti) result in an overestimation of the EWs (bottom panels of Figure~\ref{fig:fittestsfhdust}). This can be severe ($>50\%$) for $\ebmvs>0.25$ (assuming an SMC model).

In summary, we find that model assumption affect the fitting results only mildly (less than $20\%$ in most cases). Their impact on the results of dusty galaxies ($\ebmvs > 0.3$) is larger. These findings are very weakly dependent on age. This mainly reflects the independence of the optical continuum of young ($<1\,{\rm Gyr}$) galaxies on stellar population parameters such as metallicity, SFH, and age (see also Section~\ref{sec:emlinfitmethod}). On the other hand, significant dust attenuation affects the results.
In the following, we take these biases into account by deriving the \halpha~properties the galaxies for different models with diverging properties as listed in Table~\ref{tab:fitinput}. This will define a conservative range where we expect the \halpha~emission properties to lie.

\subsection{Biases due to sensitivity limits}

Next, we explore the effect of flux limits (or S/N) on the \halpha~emission measurements.
For this, we create a similar set of mock galaxies at $z\sim4.5$ as described in Section~\ref{sec:modeldependencetest}. We assume a range in \halpha~EWs (275, 550, 1100, and 2750${\rm \AA}$ rest-frame) and dust attenuation ($\ebmvs = $0, 0.2, 0.5, assuming a Calzetti parameterization). We fix the stellar population age to $300\,{\rm Myrs}$ and use a constant SFH with solar metallicity. Subsequently, we normalize the SED to different Spitzer $4.5\,{\rm \mu m}$ magnitudes ranging from 25 to 21 AB according to the observed range (see Figure~\ref{fig:sample0}) and add Gaussian noise to the extracted model colors that are consistent with the limits of the \textit{SPLASH} survey ($25.5\,{\rm AB}$ at $3\sigma$ at $3.6\,{\rm \mu m}$ and $4.5\,{\rm \mu m}$).
The \halpha~emission properties of the mock galaxies are measured from their colors (now including photometric scatter). In order to investigate biases due to S/N only, we assume for the fitting the same stellar population model as used for creating the mock galaxies. We also assume that we know the input age and dust PDF.

Figure~\ref{fig:fittestsSN} shows the result of this test by comparing the difference in input and measured \halpha~EW as a function of input \halpha~EW. Each of the panels shows a different dust attenuation and the colors of the points denote the different magnitudes at $4.5\,{\rm \mu m}$ (which is not contaminated by \halpha) as well as the corresponding S/N in parenthesis.
The amount of bias depends largely on the dust attenuation.
For dust-free galaxies, no significant biases are expected down to $25^{\rm th}$ magnitude, corresponding to S/N~$\sim 5$.
For dust-poor galaxies ($\ebmvs\sim0.2$), no significant biases are expected down to magnitude of $24\,{\rm AB}$. For $25^{\rm th}$ magnitude, however, we would expect that low EWs ($<500\,{\rm \AA}$) are overestimated by a factor of $2-5$ ($0.3-0.7\,{\rm dex}$).
For dust-rich galaxies ($\ebmvs\sim0.5$), the biases are more severe, mostly resulting in an overestimation of EWs.

In order to assess the impact of these biases on our measurements, it is important to look at the parameter space of the observations. As shown in Figure~\ref{fig:sample0}, there is a strong trend of dust attenuation with stellar mass and therefore $4.5\,{\rm \mu m}$ magnitude. This is expected as more massive galaxies tend to be more dust-rich. Specifically, in our final luminosity selected sample, galaxies around $\ebmvs\sim0.2$ ($0.5$) are generally brighter than $24^{\rm th}$ ($22^{\rm nd}$) magnitude. Comparing these values to Figure~\ref{fig:fittestsSN}, we do not expect any severe biases in our \halpha~emission measurements beyond a factor of two in our final sample of $\logm > 9.7$ galaxies.




\bibliographystyle{aasjournal}
\bibliography{bibli.bib}

%\begin{thebibliography}{}
%\bibitem[Astropy Collaboration et al.(2013)]{2013A&A...558A..33A} Astropy Collaboration, Robitaille, T.~P., Tollerud, E.~J., et al.\ 2013, \aap, 558, A33 
%\end{thebibliography}



\end{document}